\documentclass[nofootinbib,amsfonts,prd,aps]{revtex4}
\usepackage{subfig}
\usepackage{amsmath,amssymb,amsfonts,amsthm,mathrsfs}
\usepackage{stackrel}
\usepackage{graphicx}
\usepackage{nameref}

\begin{document}

\title{Quantum fluctuating geometries and the information paradox}

\author{Rodrigo Eyheralde$^{1}$, Miguel Campiglia$^{1}$, Rodolfo Gambini$^{1}$, 
Jorge Pullin$^{2}$}
\affiliation {
1. Instituto de F\'{\i}sica, Facultad de Ciencias, 
Igu\'a 4225, esq. Mataojo, 11400 Montevideo, Uruguay. \\
2. Department of Physics and Astronomy, Louisiana State University,
Baton Rouge, LA 70803-4001}

\begin{abstract}
  We study Hawking radiation on the quantum space-time of a collapsing
  null shell. We use the geometric optics approximation as in
  Hawking's original papers to treat the radiation. The quantum space-time is constructed by superposing the classical geometries associated with collapsing
  shells with uncertainty in their position and mass. We
  show that there are departures from thermality in the radiation even
  though we are not considering back reaction. One recovers the usual profile for the
  Hawking radiation as a function of frequency in the limit where the
  space-time is classical.  However, when quantum corrections are taken
  into account, the profile of the Hawking radiation as a function of
  time contains information about the initial state of the collapsing
  shell. More work will be needed to determine if all the information
  can be recovered. The calculations show that non-trivial quantum
  effects can occur in regions of low curvature when horizons are
  involved, as for instance advocated in the firewall scenario.
\end{abstract}

\maketitle

\section{Introduction}

Black hole evaporation is perhaps the salient problem of fundamental
physics nowadays, since it tests gravity, quantum field theory and
thermodynamics in their full regimes. Hawking's calculation showing
that black holes radiate a thermal spectrum initiated the study of
this phenomenon. However, the calculation assumes a fixed given
space-time, whereas it is expected that the black hole loses mass
through the radiation and eventually evaporates completely. Associated
with the evaporation process is the issue of loss of information,
whatever memory of what formed the black hole is lost as it evaporates
in a thermal state characterized by only one number, its
temperature. Having a model calculation that follows the formation of
a black hole and its evaporation including quantum effects would be
very useful to gain insights into the process. Here we would like to
present such a model. We will consider the collapse of a null
shell. The associated space-time is very simple: it is Schwarzschild
outside the shell and flat space-time inside. We will consider a
quantum evolution of the shell with uncertainty in its position and
momentum and we will superpose the corresponding space-times to
construct a quantum space-time. On it we will study the emission of
Hawking radiation in the geometric optics approximation. We will see
that in the classical limit one recovers ordinary Hawking
radiation. However, when quantum fluctuations of the collapsing shell
are taken into account we will see that non vanishing off-diagonal
terms appear in the density matrix representing the field. The
correlations and the resulting profile of particle emission are
modulated with information about the initial quantum state of the
shell, showing that information can be retrieved. At the moment we do
not know for sure if all information is retrieved.

The model we will consider is motivated in previous studies of the
collapse of a shell \cite{lwf,hajicek,shellqg}. In all these, an
important role is played by the fact that that there are two conjugate
Dirac observables. One of them is the ADM mass of the shell. The other
is related to the position along scri minus from which the shell was
sent inwards. These studies are of importance because they show that
the quantization of the correct Dirac observables for the problem lead
to a different scenario than those considered in the past using
other reduced models of the fluctuating horizon of the shell (see for
instance \cite{medvedetal}).

The organization of this paper is as follow. In the next section we
review the calculation of the radiation with a background given by a
classical collapsing shell for late times in the geometric optics
approximation, mostly to fix notation to be used in the rest of the
work.  In section 3 we will remove the late time approximation
providing an expression of the radiation of the shell for all times.
We will also derive a closed expression for the distribution of
radiation as a function of the position of the detector on scri
plus. We will show that when the shell approaches the horizon the
usual thermal radiation is recovered.  
We will see that the use of the complete expression for all times is
useful when one considers the case of fluctuating horizons in the
early (non-thermal) phases of the radiation prior to the formation of
a horizon. This element had been missed in previous calculations that
tried to incorporate such effects. In section 4 we will consider a
quantum shell and the radiation it produces, we will proceed in two
stages. First we will compute the expectation value of Bogoliubov
coefficients. This will allow to explain in a simple case the
technique that shall be used. However, the calculation of the number
of particles produced requires the expectation value of a product of
Bogoliubov coefficients. In section 5 
we consider the calculation of the density
matrix in terms of the product of Bogoliubov operators and show that
the radiation profile reproduces the usual thermal spectrum for the
diagonal elements of the density matrix, but with some departures due
to the fluctuations in the mass of the shell.
In section 6 we will show that it differs significantly
from the product of the expectation values, particularly in the late
stages of the process.   In section 7 we will
analyze coherences that vanish in the classical case
and show they are non-vanishing and that allow information from the
initial state of the shell to be retrieved. We end with a summary and outlook.

\section{Radiation of a collapsing classical shell}

Here we reproduce well known results
\cite{h74} for the late time radiation of a
collapsing classical shell in a certain amount of
detail since we will use them later on. The metric of the
space-time is given by
\begin{equation}
ds^{2}=-\left(1-\frac{2M\theta(v-v_{s})}{r}\right)dv^{2}+2dvdr+r^{2}d\Omega^{2},
\end{equation}
where $v_s$ represents the position of the shell (in ingoing
Eddington--Finkelstein coordinates) and $M$ its mass 
\footnote{
The parameters $v_s$ and $M$ are canonically conjugate variables in a Hamiltonian treatment of the system \cite{lwf}. They will be promoted to quantum operators in section IV.}.
Throughout this paper we will be  working in the geometric
optics approximation (i.e. large frequencies).
In this
geometry, light rays that leave $I^{-}$ with coordinate $v$ less than 
$v_{0}=v_{s}-4M$ can escape to $I^{+}$ and the rest are trapped in the
black hole that forms. Therefore $v=v_0$ defines the position of the
event horizon. We will use that a light ray departing from $I^-$
with $v<v_0$ reaches $I^+$ at an outgoing Eddington--Finkelstein
coordinate $u$ given by 
\begin{equation}
u(v)=v-4M \ln \left(\frac{v_{0}-v}{4M_{0}}\right)\label{eq:u(v)},
\end{equation}
where $M_0$ is an arbitrary parameter that is usually chosen as
$M_0=M$, stemming from the definition of the tortoise coordinate which
involves a constant of integration. 
  \begin{figure}
\includegraphics[height=10.5cm]{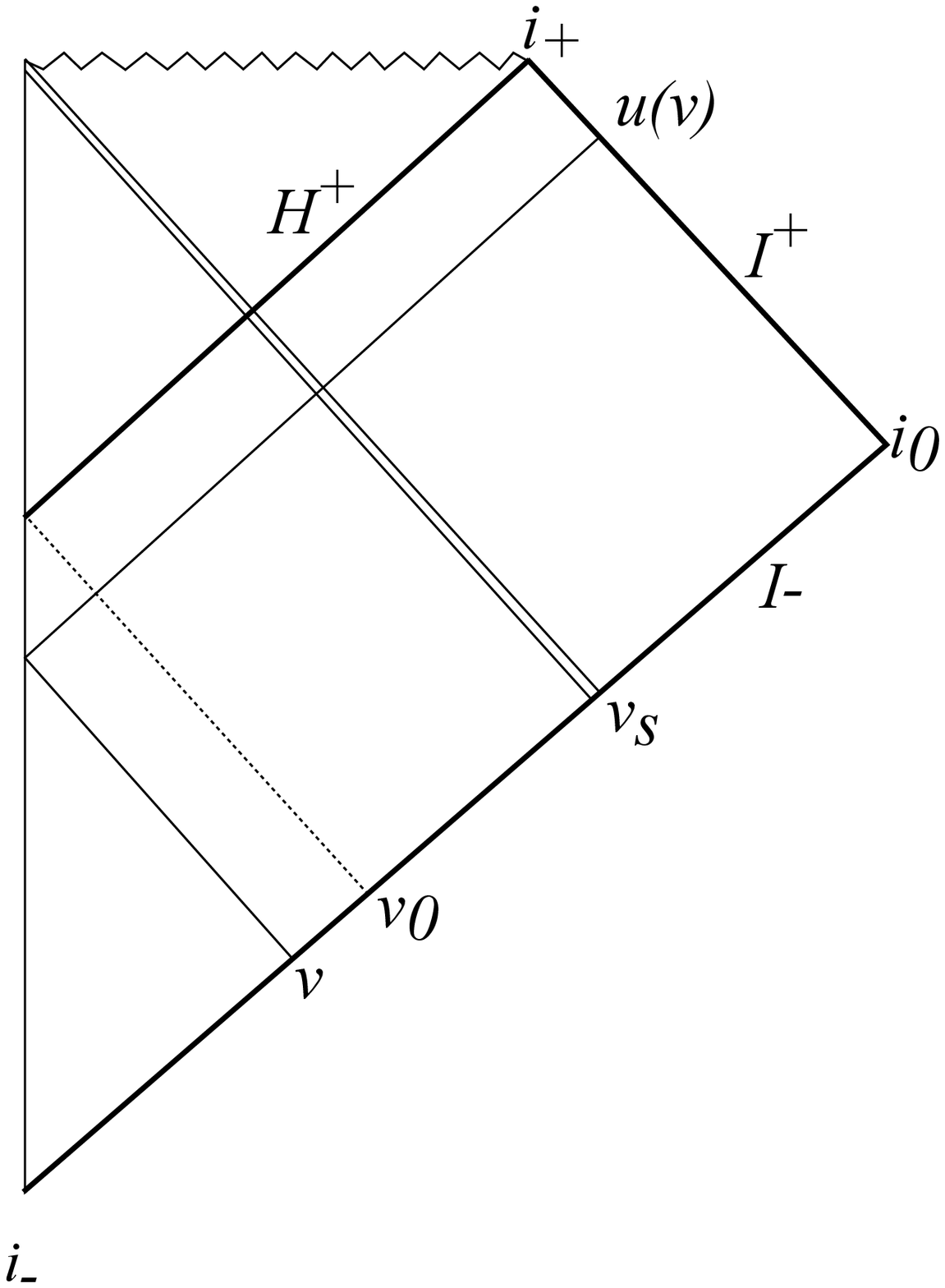}
\caption{The Penrose diagram of a classical collapsing shell. $v_s$ indicates
  the position at scri minus from which the shell is sent in. Light
  rays sent in to the left of $v_0$ make it to scri plus, whereas rays
  sent in to the right of $v_0$ get trapped in the black hole.}
\end{figure}
On the above metric we would like to study Hawking radiation
corresponding to a scalar field. We consider the ``in'' vacuum
associated with the mode expansion $\psi_{lm\omega'}$. The asymptotic
form of the modes in $I^-$ is given by,
\[
\psi_{lm\omega'}(r,v,\theta,\phi)=\frac{e^{-i\omega'v}}{4\pi r\sqrt{\omega'}}Y_{lm}(\theta,\phi),
\]
and the ``out'' vacuum corresponding to modes $\chi_{lm\omega}$ with
asymptotic form in $I^+$ given by 
\[
\chi_{lm\omega}(r,u,\theta,\phi)=\frac{e^{-i\omega u}}{4\pi r\sqrt{\omega}}Y_{lm}(\theta,\phi).
\]
The geometric optics approximation consists of mapping the modes
$\chi_{lm\omega}$ into $I^-$ as
\[
\frac{e^{-i\omega u(v)}}{4\pi r\sqrt{\omega}}Y_{lm}(\theta,\phi),
\]
where $u\left(v\right)$ is determined by the path of the light rays
that emanate from $I^-$ at time $v$ and arrive in $I^+$ at $u(v)$. 

The Bogoliubov coefficients are given by the 
Klein-Gordon inner products,
\[
\alpha_{\omega\omega'}=\left\langle \chi_{lm\omega},\psi_{lm\omega'}\right\rangle, 
\]
\[
\beta_{\omega\omega'}=-\left\langle \chi_{lm\omega},\psi_{lm\omega'}^{*}\right\rangle. 
\]
They can be computed in the geometric optics approximation projecting
the out modes in $I^-$ and substituting the expression for $u(v)$. Focusing on the beta coefficient we get,
\begin{equation}
\beta_{\omega\omega'}=-\frac{1}{2\pi}\sqrt{\frac{\omega'}{\omega}}
{\int_{-\infty}^{v_0}}dve^{-i\omega\left[v-4M\ln\left(\frac{v_{0}-v}{4M_{0}}\right)\right]- i\omega'v}.\label{refA}
\end{equation}
Since we are considering modes that are not normalizable one in
general will get divergences. This can be dealt with by considering
wave-packets localized in both frequency and time. For example,
\begin{equation}
\chi_{lmn\omega_{j}}=\frac{1}{\sqrt{\epsilon}}{\int_{j\epsilon}^{\left(j+1\right)\epsilon}}d\omega e^{u_n\omega i}\chi_{lm\omega}\label{eq:paquete_def},
\end{equation}
constitute an orthonormal countable complete basis of packets centered
in time $u_n=\frac{2\pi n}{\epsilon}$, and in frequency $\omega_{j}=\left(j+\frac{1}{2}\right)\epsilon$.

The original Hawking calculation assumes that the rays depart just
before the formation of the horizon and arrive at $I^+$ at late
times. In that case one can approximate,
\[
u(v)=v-4M\ln\left(\frac{v_{0}-v}{4M_{0}}\right)\approx v_{0}-4M\ln\left(\frac{v_{0}-v}{4M_{0}}\right).
\]
Defining a new integration variable 
$x=\frac{v_{0}-v}{4M_{0}}$ one gets
\begin{equation}
\beta_{\omega\omega'}=-\frac{4M_{0}}{2\pi}\sqrt{\frac{\omega'}{\omega}}\underset{\epsilon\to0}{\lim}{\int_0^\infty}dxe^{-i\omega\left[v_{0}-4M\ln\left(x\right)\right]- i\omega'\left(v_{0}-4M_{0}x\right)}e^{-\epsilon x}\label{eq:coef_bogol_1},
\end{equation}
where the last factor was added to make the integral convergent since
we have used plane waves instead of localized packets as the basis of
modes, following Hawking's original derivation. Using the identity 
\begin{equation}
{\int_0^\infty}dxe^{a\ln\left(x\right)}e^{-bx}=e^{-(1+a)\ln\left(b\right)}\Gamma\left(1+a\right),\quad Re(b)>0\label{eq:propiedad_int_gamma},
\end{equation}
and the usual prescription for the logarithm of a complex variable we can take the limit and get
\begin{equation}
\beta_{\omega\omega'}=-\frac{i}{2\pi}\frac{e^{-i\left(\omega+\omega'\right)v_{0}}}{\sqrt{\omega\omega'}}e^{-2\pi M\omega}\Gamma\left(1+4M\omega i\right)e^{-4M\omega i\ln(4M_{0}\omega')}\label{eq:coef_bogol2}.
\end{equation}
Now from the Bogoliubov coefficients we can calculate the expectation value of the number of particles per unit frequency detected at scri using
\[
\left\langle N^{H}_{\omega}\right\rangle ={\int}_0^\infty d\omega'\beta_{\omega\omega'}\beta_{\omega\omega'}^{*}=\frac{1}{4\pi^{2}\omega}e^{-4\pi M\omega}\left|\Gamma\left(1+4M\omega i\right)\right|^{2}{\int_0^\infty}d\omega'\frac{1}{\omega'},
\]
where we added the superscript ``$H$'' to indicate this is the
calculation originally carried out by Hawking.

The pre-factor is computed using the identity
\[
\Gamma\left(1+z\right)\Gamma\left(1-z\right)=\frac{z\pi}{\sin\left(z\pi\right)},
\]
with $z=4M\omega i$, which leads to,
\[
\left|\Gamma\left(1+4M\omega i\right)\right|^{2}=\frac{8M\pi\omega}{e^{+4M\omega\pi}-e^{-4M\omega\pi}}.
\]
To handle the divergent integral we note that 
\[
{\int_0^\infty}d\omega'\frac{1}{\omega'}=\underset{\alpha\to0}{\lim}
{\int_0^\infty}d\omega'\frac{1}{\omega'}e^{i4M\alpha\ln(\omega')}=\left[\begin{array}{l}
y=\ln\left(\omega'\right)\\
dy=\frac{d\omega'}{\omega'}
\end{array}\right]=
\]
\[
=\underset{\alpha\to0}{\lim}{\int_0^\infty}dye^{i4M\alpha y}=\frac{1}{4M}\delta\left(0\right).
\]

Therefore,
\begin{equation}
\langle N_\omega^H\rangle=\frac{1}{e^{8M\omega\pi}-1}\frac{4M}{2\pi}
{\int_0^\infty}d\omega'\frac{1}{\omega'}=\frac{1}{e^{8M\omega\pi}-1}\delta\left(0\right)\label{eq:N_omega_1}.
\end{equation}

Again, the results is infinite because we considered plane waves. The
time of arrival has infinite uncertainty and we are therefore adding
up all the particles generated for an infinite amount of time. To deal
with this we can consider wave-packets centered in time 
$u_n$ and frequency $\omega_{j}$ for
which the Bogoliubov coefficients are,
\[
\beta_{\omega_{j}\omega'}=\frac{1}{\sqrt{\epsilon}}
{\int_{j\epsilon}^{{\left(j+1\right)\epsilon}}}d\omega e^{u_n\omega i}\beta_{\omega\omega'}.
\]

We start computing the density matrix
\[
\rho^H_{\omega_1,\omega_2}={\int_0^\infty}d\omega'\beta_{\omega_{1}\omega'}\beta_{\omega_{2}\omega'}^{*}=\frac{1}{4\pi^{2}\sqrt{\omega_{1}\omega_{2}}}e^{-i\left(\omega_{1}-\omega_{2}\right)v_{0}}e^{-2\pi M\left(\omega_{1}+\omega_{2}\right)}\Gamma\left(1+4M\omega_{1}i\right)\Gamma\left(1-4M\omega_{2}i\right)\times
\]
\[
\times{\int_0^\infty}d\omega'\frac{1}{\omega'}e^{-4M\left(\omega_{1}-\omega_{2}\right)\ln\left(4M_{0}\omega'\right)}=\left[\begin{array}{l}
y=\ln\left(4M_{0}\omega'\right)\\
dy=\frac{d\omega'}{\omega'}
\end{array}\right]=
\]
\[
=\frac{1}{4\pi^{2}\sqrt{\omega_{1}\omega_{2}}}e^{-i\left(\omega_{1}-\omega_{2}\right)v_{0}}e^{-2\pi M\left(\omega_{1}+\omega_{2}\right)}\Gamma\left(1+4M\omega_{1}i\right)\Gamma\left(1-4M\omega_{2}i\right){\int_{-\infty}^\infty}dye^{-4M\left(\omega_{1}-\omega_{2}\right)y}=
\]
\begin{equation}
=\frac{1}{4\pi^{2}\omega_{1}}e^{-4\pi M\omega_{1}}\left|\Gamma\left(1+4M\omega_{1}i\right)\right|^{2}2\pi\delta\left(4M\left(\omega_{1}-\omega_{2}\right)\right)=\frac{1}{e^{8M\omega_{1}\pi}-1}\delta\left(\omega_{1}-\omega_{2}\right).\label{eq:density_matrix_H}
\end{equation}

Therefore, 
\begin{equation}
\langle N_{\omega_j}^H\rangle={\int_0^\infty}d\omega'\beta_{\omega_{j}\omega'}\beta_{\omega_{j}\omega'}^{*}=
\frac{1}{\epsilon}{\int\int_{j\epsilon}^{\left(j+1\right)\epsilon}}d\omega_{1}d\omega_{2}
e^{u_n\left(\omega_{1}-\omega_{2}\right)i}\rho^H_{\omega_1,\omega_2}
=\frac{1}{\epsilon}{\int_{j\epsilon}^{\left(j+1\right)\epsilon}}\frac{1}{e^{8M\omega_{1}\pi}-1}d\omega_{1}\sim\frac{1}{e^{8M\omega_{j}\pi}-1}\label{eq:N_omega_2},
\end{equation}
which is the standard result for the Hawking radiation spectrum.

\section{Calculation without approximating $u(v)$} 

We will carry out the computation of the Bogoliubov
coefficients using the exact expression for $u(v)$. This will be of
importance for the case with quantum fluctuations. This is because if
one looks at the expression of the time of arrival,
\begin{equation}
u(v)=v-4M\ln\left(\frac{v_{0}-v}{4M_{0}}\right),
\end{equation}
when one has quantum fluctuations, even close to the horizon, the
second term is not necessarily very large. For instance, if one
considers fluctuations of Planck length size and a Solar sized black
hole, it is around $100M$. Therefore it is not warranted to neglect
the first term as we did in the previous section. In this section we
will not consider quantum fluctuations yet. However, using the exact
expression allows to compute the radiation emitted by a shell 
far away from the horizon.

Starting with the
expression:
\[
\beta_{\omega\omega'}=-\frac{1}{2\pi}\sqrt{\frac{\omega'}{\omega}}
{\int_{-\infty}^{v_{0}}}dve^{i4M\omega\ln\left(\frac{v_{0}-v}{4M_{0}}\right)- i\omega'v}e^{-i\omega v},
\]
we change variables to 
 $x=\frac{v_{0}-v}{4M_{0}}$ and introduce a regulator $e^{-\epsilon
   x}$. We get,
\begin{equation}
\beta_{\omega\omega'}=-\frac{4M_{0}}{2\pi}\sqrt{\frac{\omega'}{\omega}}e^{-i\left(\omega+\omega'\right)v_{0}}\underset{\epsilon\to0}{\lim}{\int_0^\infty}dxe^{i4M\omega \ln\left(x\right)}e^{-\left(\epsilon-i\left[\omega+\omega'\right]4M_{0}\right)x}.
\label{betaclasico}
\end{equation}
For $\omega\ll\omega'$ we recover Hawking's original
calculation. However, we can continue without approximating. Using
again (\ref{eq:propiedad_int_gamma}) we get,
\[
\beta_{\omega\omega'}=-\frac{4M_{0}}{2\pi}\sqrt{\frac{\omega'}{\omega}}e^{-i\left(\omega+\omega'\right)v_{0}}\Gamma(1+4M\omega i)\underset{\epsilon\to0}{\lim}\;e^{-(1+4M\omega i)\ln\left(\epsilon-i\left[\omega+\omega'\right]4M_{0}\right)}.
\]
And taking the limit,
\begin{equation}
\beta_{\omega\omega'}=-\frac{i}{2\pi}\frac{1}{\omega'+\omega}\sqrt{\frac{\omega'}{\omega}}e^{-i\left(\omega+\omega'\right)v_{0}}\Gamma(1+4M\omega i)e^{-2\pi M\omega}e^{4M\omega i\ln\left(4M_{0}\left[\omega'+\omega\right]\right)}\label{eq:coef_bogol3}.
\end{equation}
To compare with Hawking's calculation we first compute
\[
\langle N_\omega^{CS}\rangle={\int_0^\infty}d\omega'\beta_{\omega\omega'}\beta_{\omega\omega'}^{*}=\frac{1}{4\pi^{2}}\frac{1}{\omega}\left|\Gamma(1+4M\omega i)\right|^{2}e^{-4\pi M\omega}{\int_0^\infty}d\omega'\frac{\omega'}{\left(\omega'+\omega\right)^{2}},
\]
where the superscript ``$CS$'' stands for classical shell. 
The difference with the calculation in the previous section is
the argument of the last integral with no divergence in $\omega'=0$.

We can formally compute the divergent integral using the change
of variable  $y=\ln\left(\omega'+\omega\right)$. We get,
\[
{\int_0^\infty}d\omega'\frac{\omega'}{\left(\omega'+\omega\right)^{2}}=
{\int_{\ln(\omega)}^{\infty}}dye^{-y}\left(e^{y}-\omega\right)=
{\int_{\ln(\omega)}^{\infty}}dy-1=
\]
\[
=\left.{\int_{\ln(\omega)}^{\infty}}dye^{i4M\alpha
    y}\right|_{\alpha=0}-1=\left.
{\int_0^\infty}dye^{i4M\alpha
  y}e^{i4M\alpha\ln(\omega)}\right|_{\alpha=0}-1=\frac{1}{4M}\left(\pi\delta\left(0\right)+{\rm
  p.v.}\left(\frac{i}{0}\right)\right)-1,
\]
with ${\rm p.v.}$ the principal value. 
Therefore,
\begin{equation}
\langle N_\omega^{CS}\rangle=\frac{1}{e^{8M\omega\pi}-1}\frac{4M}{2\pi}
{\int_0^\infty}d\omega'\frac{\omega'}{\left(\omega'+\omega\right)^2}=\frac{1}{e^{8M\omega\pi}-1}\left[\left(\frac{\delta\left(0\right)}{2}+{\rm
      p.v.}\left(\frac{i}{2\pi0}\right)\right)-\frac{2M}{\pi}\right]\label{eq:N_omega3}.
\end{equation}

This is an infinite result but it looks different from Hawking's. To deal  with the
infinities it is necessary to compute $\langle {N_{\omega_j}}^{CS}\rangle$
for a wave-packet of frequency
 $\omega_{j}$. We start by computing the density matrix:
\[
\rho^{CS}_{\omega_1,\omega_2}={\int_0^\infty}d\omega'\beta_{\omega_{1}\omega'}\beta_{\omega_{2}\omega'}^{*}=\frac{1}{4\pi^{2}\sqrt{\omega_{1}\omega_{2}}}e^{-i\left(\omega_{1}-\omega_{2}\right)v_{0}}\Gamma\left(1+4M\omega_{1}i\right)\Gamma\left(1-4M\omega_{2}i\right)e^{-2\pi M\left[\omega_{1}+\omega_{2}\right]}\times
\]
\begin{equation}
\times{\int_0^\infty}d\omega'\frac{\omega'}{\left(\omega'+\omega_{1}\right)\left(\omega'+\omega_{2}\right)}e^{-4Mi\left[\omega_{1}\ln\left(4M_{0}\left[\omega'+\omega_{1}\right]\right)-\omega_{2}\ln\left(4M_{0}\left[\omega'+\omega_{2}\right]\right)\right]}.\label{casoclasico}
\end{equation}

Since the packet is centered in 
$\omega_{j}$ with width $\epsilon\ll\omega_j$ we introduce $\Delta\omega=\omega_{2}-\omega_{1}+$ and $\bar{\omega}=\frac{\omega_1+\omega_2}{2}$. As a consequence, the last
integral takes the form,
\[
{\int_0^\infty}d\omega'\frac{\omega'e^{-4Mi\left[\left(\bar{\omega}-\frac{\Delta\omega}{2}\right)\ln\left(4M_{0}\left[\omega'+\bar{\omega}-\frac{\Delta\omega}{2}\right]\right)-\left(\bar{\omega}+\frac{\Delta\omega}{2}\right)\ln\left(4M_{0}\left[\omega'+\bar{\omega}+\frac{\Delta\omega}{2}\right]\right)\right]}}{\left(\omega'+\bar{\omega}\right)^2-\left(\frac{\Delta\omega}{2}\right)^2}=
\]
\[
={\int_0^\infty}d\omega'\frac{\omega'e^{4Mi\Delta\omega \ln\left(4M_{0}\left[\omega'+\bar{\omega}\right]\right)}}{\left(\omega'+\bar{\omega}\right)^{2}}+O\left(\Delta\omega\right),
\]
where we have not expanded the exponential 
$e^{4Mi\Delta\omega \ln\left(4M_{0}\left[\omega'+\bar{\omega}\right]\right)}$
since it controls the divergent part of the integral when $\Delta\omega\to 0$. Changing
variable to $y=\ln\left(4M_{0}\left[\omega'+\bar{\omega}\right]\right)$ the integral becomes,
\[
{\int_{\ln\left(4M_{0}\bar{\omega}\right)}^{\infty}}dy\left(1-4M_{0}\bar{\omega}e^{-y}\right)e^{4Mi\Delta\omega y}+O\left(\Delta\omega\right)=
\]
\[
={\int_0^\infty}dye^{4Mi\Delta\omega y}e^{4Mi\Delta\omega\ln\left(4M_{0}\bar{\omega}\right)}+\frac{e^{4Mi\Delta\omega\ln\left(4M_{0}\bar{\omega}\right)}}{-1+4M_{0}\Delta\omega i}+O\left(\Delta\omega\right)=
\]
\[
=\left[\pi\delta\left(4M\Delta\omega\right)+{\rm p.v.}\left(\frac{i}{4M\Delta\omega}\right)\right]e^{4Mi\Delta\omega\ln\left(4M_{0}\bar{\omega}\right)}+O\left(\Delta\omega^0\right).
\]
So, the divergent part of the density matrix when $\Delta\omega\to0$ is
\[
\rho^{CS}_{\omega_1,\omega_2}\sim\frac{1}{4\pi^{2}\bar{\omega}}e^{i\Delta\omega
  v_{0}}\left|\Gamma\left(1+4M\bar{\omega}i\right)\right|^{2}e^{-4\pi
  M\bar{\omega}}\left[\pi\delta\left(4M\Delta\omega\right)+{\rm p.v.}\left(\frac{i}{4M\Delta\omega}\right)\right]e^{4Mi\Delta\omega\ln\left(4M_{0}\bar{\omega}\right)}.
\]
\begin{equation}
=\frac{2M}{\pi}\frac{e^{4Mi\Delta\omega\ln\left(4M_{0}\bar{\omega}\right)}}{e^{8M\omega_{j}\pi}-1}\left[\pi\delta\left(4M\Delta\omega\right)+{\rm p.v.}\left(\frac{i}{4M\Delta\omega}\right)\right].\label{eq:density_matrix_CS}
\end{equation}
We proceed to compute $\langle N_{\omega_j}^{CS}\rangle$
by integrating both Bogoliubov coefficients in an interval around 
 $\omega_{j}$ using the approximation that factors depending on $\bar{\omega}$ are constant since the interval of integration is very small as it ranges between $\omega_j\pm\frac{\epsilon-\vert\Delta\omega\vert}{\epsilon}$,
\[
\langle N_{\omega_j}^{CS}\rangle=
\frac{1}{\epsilon}{\int_{j\epsilon}^{\left(j+1\right)\epsilon}\int_{j\epsilon}^{\left(j+1\right)\epsilon}}d\omega_{1}d\omega_{2}e^{u_n\Delta\omega i}\rho^{CS}_{\omega_1,\omega_2}\sim
\]
\[
\sim\frac{1}{4\pi^{2}\omega_{j}}\frac{1}{\epsilon}\left|\Gamma\left(1+4M\omega_{j}i\right)\right|^{2}e^{-4\pi M\omega_{j}}{\int_{j\epsilon}^{\left(j+1\right)\epsilon}\int_{j\epsilon}^{\left(j+1\right)\epsilon}}d\omega_{1}d\omega_{2}e^{-\frac{2\pi n}{\epsilon}\Delta\omega i}e^{i\Delta\omega v_{0}}\times
\]
\[
\times\frac{1}{4M}\left[\pi\delta\left(\Delta\omega\right)+{\rm p.v.}\left(\frac{ie^{4M\Delta\omega\ln\left(4M_{0}\bar{\omega}\right)i}}{\Delta\omega}\right)\right]=
\]
\[
\sim\frac{1}{2\pi\epsilon}\frac{1}{e^{8M\omega_{j}\pi}-1}{\int_{j\epsilon}^{\left(j+1\right)\epsilon}\int_{j\epsilon}^{\left(j+1\right)\epsilon}}d\omega_{1}d\omega_{2}e^{-\left[u_n-v_{0}-4M\ln\left(4M_{0}\bar{\omega}\right)\right]\Delta\omega
  i}\left[\pi\delta\left(\Delta\omega\right)+{\rm p.v.}\left(\frac{i}{\Delta\omega}\right)\right].
\]
Changing variables to $\bar{\omega}$ and $\Delta\omega$ we get,
\[
\langle N_{\omega_j}^{CS}\rangle\sim\frac{1}{e^{8M\omega_{j}\pi}-1}\left[\frac{1}{2}+\frac{i}{2\pi\epsilon}
{\int_{-\epsilon}^{\epsilon}}d\left(\Delta\omega\right) {\rm p.v.}\left(\frac{1}{\Delta\omega}\right)
{\int_{\omega_{j}-\frac{\epsilon-\left|\Delta\omega\right|}{2}}^{\omega_{j}+\frac{\epsilon-\left|\Delta\omega\right|}{2}}}e^{-\left[u_n-v_{0}-4M\ln\left(4M_{0}\bar{\omega}\right)\right]\Delta\omega i}d\bar{\omega}\right]\sim
\]
\[
\sim\frac{1}{e^{8M\omega_{j}\pi}-1}\left[\frac{1}{2}+\frac{i}{2\pi}
{\int_{-\epsilon}^{\epsilon}}d\left(\Delta\omega\right) {\rm p.v.}\left(\frac{\epsilon-\left|\Delta\omega\right|}{\epsilon\Delta\omega}\right)e^{-\alpha\Delta\omega i}\right],
\]
where we defined  
\begin{equation}
\alpha\equiv u_n-v_{0}-4M\ln\left(4M_{0}\omega_{j}\right)\label{alpha}.
\end{equation}
Notice that there appears the indeterminate parameter $M_0$. This corresponds to the choice of origin of the affine parameter at scri plus.

A further change of variable  $t=\alpha\Delta\omega$ leads us to
\begin{equation}
\langle N_{\omega_j}^{CS}\rangle=\frac{1}{e^{8M\omega_{j}\pi}-1}\left[\frac{1}{2}+\frac{1}{\pi}{\rm Si}\left(\epsilon\alpha\right)+\frac{1}{\pi}\frac{\cos\left(\alpha\epsilon\right)-1}{\alpha\epsilon}\right]\label{eq:N_omega4}
\end{equation}
where ${\rm Si}$ is the sine integral. When 
$\epsilon\alpha\to\infty$
we have that ${\rm Si}\left(\epsilon\alpha\right)\to\frac{\pi}{2}$ and
the expression goes to
\[\langle N_{\omega_j}^{CS}\rangle\to
\frac{1}{e^{8M\omega_{j}\pi}-1}.
\]
 This happens when either  $n\to+\infty$ or $\omega_{j}\to0$. That is,
at late times or in the deep infra-red regime. 
On the
 other hand, 
when $n\to-\infty$ (a detector close to spatial infinity or very early
times) we have that  ${\rm Si}\left(\epsilon\alpha\right)\to-\frac{\pi}{2}$
and therefore
\[
\langle N_{\omega_j}^{CS}\rangle \to0.
\]

We have obtained a closed form for the spectrum of the
radiation of the classical shell along its complete trajectory. It only becomes
thermal at late times. This agrees with previous
numerical results \cite{vachaspati}. Previous efforts had differing
predictions on the thermality or not of the radiation \cite{previous}.

\section{Radiation from the collapse of a quantum shell}

\subsection{The basic quantum operators}
A reduced phase-space analysis of the shell shows that the Dirac
observables $v_s$ and $M$ are canonically conjugate variables \cite{lwf}. We thus promote them to quantum operators satisfying,
\begin{equation}
\left[\widehat{M},\widehat{v}_{s}\right]=i\hbar\widehat{I}\label{eq:conmutacio_M_v},
\end{equation}
with $\widehat{I}$ the identity operator. 
It will be more convenient to use the operator 
$\widehat{v}_{0}=\widehat{v}_{s}-4\widehat{M}$ which is also conjugate
to $\widehat{M}$. We call the expectation values of these quantities 
$\overline{M}\equiv\left\langle \hat{M}\right\rangle $
and $\overline{v}_{0}\equiv\left\langle \hat{v}_{0}\right\rangle $.

In terms of them we define the operator
\begin{equation}
\hat{u}\left(v,\widehat{v}_{0},\widehat{M}\right)=v\widehat{I}-2\left[\widehat{M}\ln\left(\frac{\widehat{v}_{0}-v\widehat{I}}{4M_{0}}\right)+\ln\left(\frac{\widehat{v}_{0}-v\widehat{I}}{4M_{0}}\right)\widehat{M}\right]\label{eq:operador_u},
\end{equation}
where $v$ is a real parameter and $M_0$ an arbitrary scale. This
operator represents the variable $u(v)$.  Given a value of the
parameter $v$ the operator $\hat{u}$ is well defined in the basis
$\left\{ v_{0}\right\} _{v_{0}\in\mathbb{R}}$ of eigenstates of
$\hat{v}_{0}$ only for eigenvalues $v_0>v$.  This is the relevant
region for the computation of Bogoliubov coefficients. It is however
convenient to provide an extension of the operator $\hat{u}$ to the
full range of $v_0$ so that one can work in the full Hilbert space of
the shell. The (quantum) Bogoliubov coefficients are independent of
such extension.
For instance, defining
the function $f_{\epsilon}(x)=\left\{ \begin{array}{l}
\ln(x),\;x\geq\epsilon\\
\ln(\epsilon),\;x<\epsilon
\end{array}\right.$ one can construct the operator 
\begin{equation}
\hat{u}_{\epsilon}\left(v,\widehat{v}_{0},\widehat{M}\right)=v\widehat{I}-2\left[\widehat{M}f_{\epsilon}\left(\frac{\widehat{v}_{0}-v\widehat{I}}{4M_{0}}\right)+f_{\epsilon}\left(\frac{\widehat{v}_{0}-v\widehat{I}}{4M_{0}}\right)\widehat{M}\right]\label{eq:operador_u_epsilon},
\end{equation}
which extends $\hat{u}$ to the full Hilbert space.  To understand the
physical meaning, we recall that for values of $v$ less than $v_0$ the
packets escape to scri, whereas for $v$ larger than $v_0$ they fall
into the black hole. The extension corresponds to considering particle
detectors that either live at scri or live on a time-like trajectory a
small distance outside the horizon. As we shall see, the Bogoliubov
coefficients will have a well-defined $\epsilon \to 0$ limit.

Next we seek for the eigenstates of $\hat{u}_{\epsilon}$. We work with
wave-functions 
$\psi\left(v_{0}\right)=\left\langle v_{0}\vert\psi\right\rangle $. The
operator $\hat{M}$ (conjugate to $\hat{v}_0$) is,
\begin{equation}
\left\langle v_{0}\vert\hat{M}\psi\right\rangle =i\hbar\frac{\partial\psi}{\partial v_{0}}\label{eq:operadorM}.
\end{equation}

The eigenstates of of $\hat{u}_\epsilon$ are given by the equation
\[
\left\langle v_{0}\vert\hat{u}_{\epsilon}\psi_{u}\right\rangle =u\psi_{u}\left(v_{0}\right),
\]
that is,
\begin{equation}
v\psi_{u}\left(v_{0}\right)-2i\hbar\frac{\partial}{\partial v_{0}}\left[f_{\epsilon}\left(\frac{v_{0}-v}{4M_{0}}\right)\psi_{u}\left(v_{0}\right)\right]-2i\hbar f_{\epsilon}\left(\frac{v_{0}-v}{4M_{0}}\right)\frac{\partial\psi}{\partial v_{0}}=u\psi_{u}\left(v_{0}\right)\label{eq:ecuaci=0000F3n_autoestados_u},
\end{equation}
\[
v\psi_{u}\left(v_{0}\right)-4i\hbar f_{\epsilon}\left(\frac{v_{0}-v}{4M_{0}}\right)\frac{\partial\psi}{\partial v_{0}}-\frac{2i\hbar}{4M_{0}}f'_{\epsilon}\left(\frac{v_{0}-v}{4M_{0}}\right)\psi_{u}\left(v_{0}\right)=u\psi_{u}\left(v_{0}\right).
\]
It is useful to make a change of variable 
$x=\frac{v_{0}-v}{4M_{0}}$ which leads to 
\[
-\frac{4i\hbar}{4M_{0}}f_{\epsilon}\left(x\right)\frac{\partial\psi}{\partial x}-\frac{2i\hbar}{4M_{0}}f'_{\epsilon}\left(x\right)\psi_{u}\left(x\right)=\left(u-v\right)\psi_{u}\left(x\right).
\]
Defining $\phi_u(x)$ by 
$\psi_{u}(x)=\frac{\phi_{u}(x)}{\sqrt{\left|f_{\epsilon}(x)\right|}}$
we get,
\[
\frac{\partial\phi_{u}}{\partial x}=\frac{iM_{0}}{\hbar}\frac{u-v}{f_{\epsilon}}\phi_{u},
\]
with general solution 
\[
\phi_{u}(x)=\phi_{0}\exp\left(\frac{iM_{0}}{\hbar}(u-v)\int\frac{ds}{f_{\epsilon}(s)}\right).
\]

\begin{figure}
\includegraphics[scale=0.7]{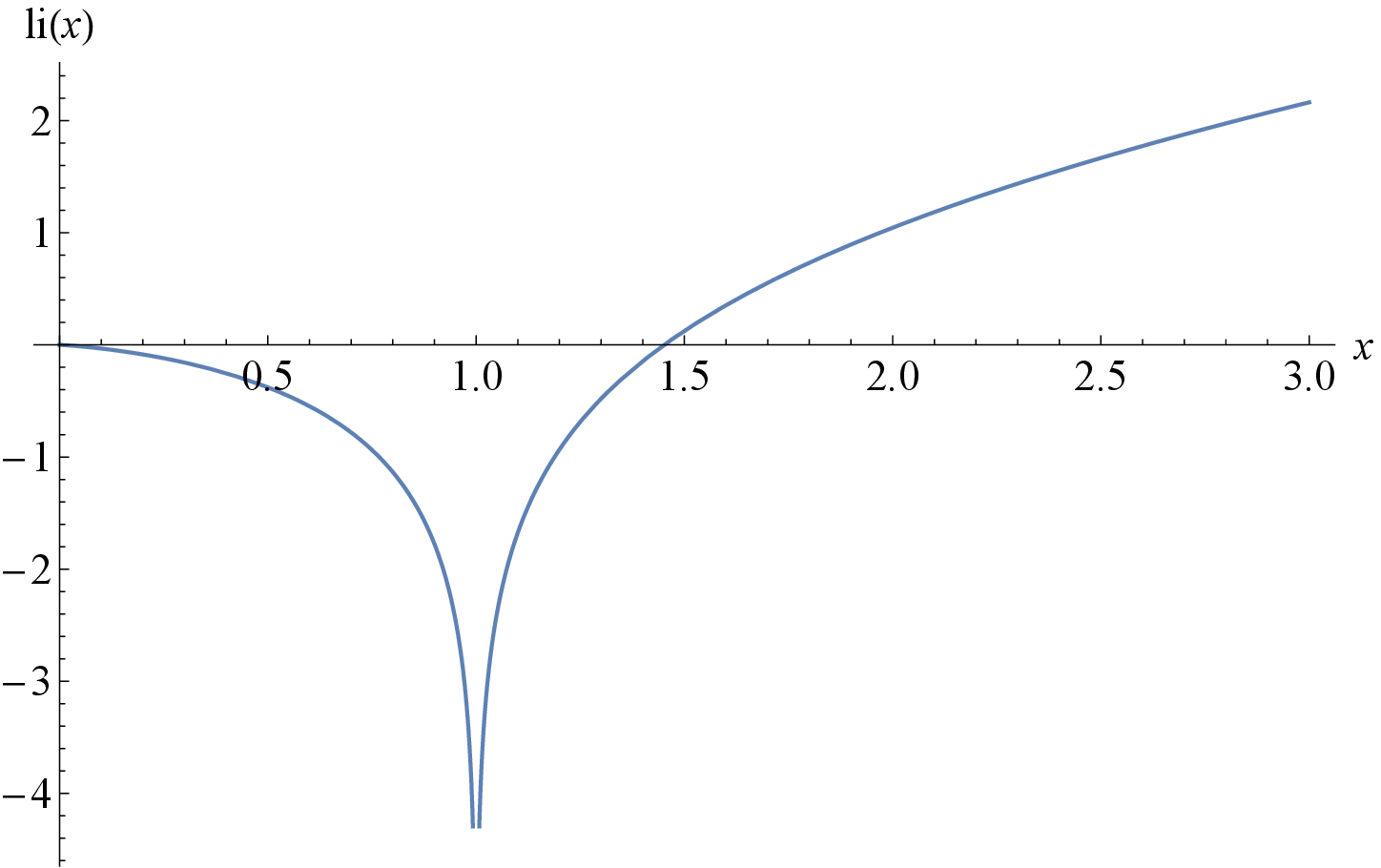}
\caption{The logarithmic integral function.}
\label{li}
\end{figure}

Substituting $f_\epsilon$ and going back to the original variables
\[
\psi_{u}(x)=\left\{ \begin{array}{l}
\frac{\psi_{0}^{I}}{\sqrt{\left|\ln(x)\right|}}\exp\left(\frac{iM_{0}}{\hbar}(u-v){\rm li}(x)\right),\quad x\geq\epsilon,\\
\frac{\psi_{0}^{II}}{\sqrt{\left|\ln(\epsilon)\right|}}\exp\left(\frac{iM_{0}}{\hbar}(u-v)\frac{x}{\ln(\epsilon)}\right),\quad x<\epsilon,
\end{array}\right.
\]
where $\phi_0, \psi_{0}^I$ and $\psi_0^{II}$ are independent, complex,
constants and 
\begin{equation}
{\rm li}(x)={\int_0^x}\frac{dt}{\ln(t)}\label{eq:Li},
\end{equation}
is the logarithmic integral, which is plotted in figure (\ref{li}).

The discontinuity of $\psi_u$ in $x=1$ introduces a degeneracy in the
eigenstates of $\hat{u}$. For each eigenvalue we can choose two
independent eigenstates,
\begin{equation}
\psi_{u}^{1}(x)=\left\{ \begin{array}{l}
\frac{1}{\sqrt{8\pi\hbar\left|\ln(\epsilon)\right|}}\exp\left(\frac{iM_{0}}{\hbar}(u-v)\frac{x-\epsilon}{\ln(\epsilon)}\right),\quad x<\epsilon\\
\frac{1}{\sqrt{8\pi\hbar\left|\ln(x)\right|}}\exp\left(\frac{iM_{0}}{\hbar}(u-v)\left[{\rm
      li}\left(x\right)-{\rm li}\left(\epsilon\right)\right]\right),\quad\epsilon\leq x<1\\
0,\quad x\geq 1
\end{array}\right.\label{eq:autoestados_de_u_1}
\end{equation}
\begin{equation}
\psi_{u}^{2}(x)=\left\{ \begin{array}{l}
0,\quad x\leq 1\\
\frac{1}{\sqrt{8\pi\hbar\left|\ln(x)\right|}}\exp\left(\frac{iM_{0}}{\hbar}(u-v)\left[{\rm
      li}\left(x\right)-{\rm li}\left(\epsilon\right)\right]\right),\quad x>1
\end{array}\right.\label{eq:autoestados_de_u_2}
\end{equation}
which we have chosen as orthonormal. We will adopt the notation 
$\left|u,J\right\rangle _{\epsilon}$ with $J=1,2$ for these states. 

\subsection{Operators associated with the Bogoliubov coefficients
  and their expectation values}

On the previously described quantum space-time we will study Hawking radiation
associated with a scalar field.  We will assume that the scalar field
sees a superposition of geometries corresponding to different masses
of the black hole. Therefore, to measure observables associated with
the field one needs to take their expectation value with respect to
the wave-function of the black hole.
In this subsection we will apply these ideas to the computation of the Bogoliubov coefficients and in the next we will extend it to compute the density matrix. We will go from the usual Bogoliubov coefficient $\beta_{\omega\omega'}$ to the 
operator $\hat{\beta}_{\omega\omega'}$. We will then compute its expectation
value on a wave-function packet associated to the black hole and
centered on the classical values
$\overline{M}$ and $\overline{v}_{0}$. We start with the expression 
(\ref{refA})  and promote it to a well defined operator
\begin{equation}
\hat{\beta}_{\omega\omega'}=-\frac{1}{2\pi}\sqrt{\frac{\omega'}{\omega}}\underset{\epsilon\to0}{\lim}{\int_{-\infty}^{+\infty}}dv\theta\left(\widehat{v}_{0}-v\widehat{I}\right)e^{-i\omega\hat{u}_{\epsilon}(v)- i\omega'v}\theta\left(\widehat{v}_{0}-v\widehat{I}\right)\label{eq:coef_bogol_cuanticos}.
\end{equation}

We then consider a state $\Psi$ associated with the black hole and
compute the expectation value,
\[
\left\langle
  \hat{\beta}\right\rangle_{\omega\omega'}=-\frac{1}{2\pi}\sqrt{\frac{\omega'}{\omega}}\underset{\epsilon\to0}{\lim}\left\langle
  \Psi\right|
{\int_{-\infty}^{+\infty}}dv{\int_{-\infty}^{+\infty}}dv_{0}\left|v_{0}\right\rangle \left\langle v_{0}\right|\theta\left(\widehat{v}_{0}-v\widehat{I}\right)e^{-i\omega\hat{u}_{\epsilon}(v)- i\omega'v}\times
\]
\[
\times\underset{J=1,2}{\sum}{\int_{-\infty}^{+\infty}}du\left|u,J\right\rangle _{\epsilon\epsilon}\left\langle u,J\right|{\int_{-\infty}^{+\infty}}dv'_{0}\left|v'_{0}\right\rangle \left\langle v'_{0}\right|\theta\left(\widehat{v}_{0}-v\widehat{I}\right)\left|\Psi\right\rangle, 
\]
where we have introduced bases of eigenstates of 
$\hat{v}_{0}$
and $\hat{u}$.

Given,
\[
\left\langle \hat{\beta}\right\rangle_{\omega\omega'}=-\frac{1}{2\pi}\sqrt{\frac{\omega'}{\omega}}\underset{\epsilon\to0}{\lim}{\int_{-\infty}^\infty\int_{-\infty}^\infty\int_{-\infty}^\infty\int_{-\infty}^\infty}dvdv_{0}dv'_{0}du\Psi^{*}(v_{0})\Psi(v'_{0})\theta\left(v_{0}-v\right)\theta\left(v'_{0}-v\right)\times
\]
\[
\times e^{-i\omega u - i\omega'v}\underset{J=1,2}{\sum}\psi_{u,J}(v_{0})\psi_{u,J}^{*}(v'_{0}),
\]
and changing variables 
$x_{1}=\frac{v_{0}-v}{4M_{0}}$
and $x_{2}=\frac{v'_{0}-v}{4M_{0}}$ 
we get 
\[
\left\langle \hat{\beta}\right\rangle_{\omega\omega'}=-\frac{\left(4M_{0}\right)^{2}}{2\pi}\sqrt{\frac{\omega'}{\omega}}\underset{\epsilon\to0}{\lim}{\int_{-\infty}^\infty}dve^{- i\omega'v}{\int_0^\infty\int_0^\infty}dx_{1}dx_{2}\Psi^{*}(4M_{0}x_{1}+v)\times
\]
\begin{equation}
\times\Psi(4M_{0}x_{2}+v){\int_{-\infty}^{+\infty}}due^{-i\omega u}\underset{J=1,2}{\sum}\psi_{u}^{J}(x_{1})\psi_{u}^{J*}(x_{2}).\label{genericexpression}
\end{equation}
The definition of the eigenstates $\psi_{u}^{I}$ reduces the integral
in 
${\int_0^\infty\int_0^\infty}dx_{1}dx_{2}$ to
\[
{\int_0^\epsilon\int_0^\epsilon}dx_{1}dx_{2}+{\int_0^\epsilon}{\int_\epsilon^1}dx_{1}dx_{2}+{\int_\epsilon^1}{\int_0^\epsilon}dx_{1}dx_{2}+{\int_\epsilon^1\int_\epsilon^1}dx_{1}dx_{2}+{\int_1^\infty\int_1^\infty}dx_{1}dx_{2}.
\]
In the appendix we show that the first 3 integrals do not contribute
in the limit $\epsilon\to 0$. Therefore the calculation reduces to,
\[
\left\langle \hat{\beta}\right\rangle_{\omega\omega'}=-\frac{\left(4M_{0}\right)^{2}}{2\pi8\pi\hbar}\sqrt{\frac{\omega'}{\omega}}\underset{\epsilon\to0}{\lim}{\int_{-\infty}^\infty}dve^{- i\omega'v}\left({\int_\epsilon^1\int_\epsilon^1}dx_{1}dx_{2}+{\int_1^\infty\int_1^\infty}dx_{1}dx_{2}\right)\Psi^{*}(4M_{0}x_{1}+v)\times
\]
\[
\times\Psi(4M_{0}x_{2}+v){\int_{-\infty}^\infty}due^{-i\omega
  u}\frac{1}{\sqrt{\left|\ln(x_{2})\right|\left|\ln(x_{1})\right|}}\exp\left(\frac{iM_{0}}{\hbar}(u-v)\left[{\rm
      li}(x_{1})-{\rm li}(x_{2})\right]\right).
\]

Computing the integral in $u$ we get,
\[
\left\langle \hat{\beta}\right\rangle_{\omega\omega'}=-\frac{\left(4M_{0}\right)^{2}}{2\pi8\pi\hbar}\sqrt{\frac{\omega'}{\omega}}\underset{\epsilon\to0}{\lim}{\int_{-\infty}^\infty}dv\,e^{- i\omega'v}\left({\int_\epsilon^1\int_\epsilon^1}dx_{1}dx_{2}+{\int_1^\infty\int_1^\infty}dx_{1}dx_{2}\right)\Psi^{*}(4M_{0}x_{1}+v)\times
\]
\[
\times\Psi(4M_{0}x_{2}+v)\frac{2\pi\delta\left(\omega-\frac{M_{0}}{\hbar}
\left[{\rm li}(x_{1})-{\rm li}(x_{2})\right]\right)}
{\sqrt{\left|\ln(x_{2})\right|\left|\ln(x_{1})\right|}}e^{-i\omega
  v}.
\]

Since ${\rm li}$ is invertible in $\left(0,1\right)$ and in $\left(1,+\infty\right)$
we can then integrate in $x_{2}$ to get 
\[
\left\langle \hat{\beta}\right\rangle_{\omega\omega'}=-\frac{2M_{0}}{\pi}\sqrt{\frac{\omega'}{\omega}}{\int_{-\infty}^\infty}dve^{- i\omega'v}\left({\int_0^1}dx_{1}+{\int_1^\infty}dx_{1}\right)\times
\]
\[
\times\Psi^{*}(4M_{0}x_{1}+v)\Psi(4M_{0}x_{2}\left(x_{1}\right)+v)\sqrt{\frac{\left|\ln(x_{2})\right|}{\left|\ln(x_{1})\right|}}e^{-i\omega v},
\]
where $x_{2}\left(x_{1}\right)={\rm li}^{-1}\left[{\rm li}\left(x_{1}\right)-\frac{\omega\hbar}{M_{0}}\right]$
and we have used that  $\partial_t{\rm li}\left(t\right)=\frac{1}{\left|\ln(t)\right|}$.
We redefine $x=x_{1}$ and 
\begin{equation}
\bar{x}_\omega(x)={\rm li}^{-1}\left[{\rm li}\left(x\right)-\frac{\omega\hbar}{M_{0}}\right]\label{eq:inversaLi}.
\end{equation}

Therefore,
\[
\left\langle \hat{\beta}\right\rangle_{\omega\omega'}=-\frac{2M_{0}}{\pi}\sqrt{\frac{\omega'}{\omega}}{\int_0^\infty}dx\sqrt{\frac{\left|\ln(\bar{x}_\omega\left(x\right))\right|}{\left|\ln(x)\right|}}{\int_{-\infty}^\infty}dve^{-i\left[\omega+\omega'\right]v}\Psi^{*}(4M_{0}x+v)\Psi(4M_{0}\bar{x}_\omega\left(x\right)+v),
\]
where we have inverted the order of the integrals for convenience of
subsequent calculations. Finally, the change of variable $s\equiv v+2M_0\left[x+\bar{x}_\omega(x)\right]$ gives us

\[
\left\langle \hat{\beta}\right\rangle_{\omega\omega'}=-\frac{2M_{0}}{\pi}\sqrt{\frac{\omega'}{\omega}}{\int_0^\infty}dx\sqrt{\frac{\left|\ln(\bar{x}_\omega\left(x\right))\right|}{\left|\ln(x)\right|}}e^{i2M_0\left[\omega+\omega'\right]\left[x+\bar{x}_\omega(x)\right]}{\int_{-\infty}^\infty}ds e^{-i\left[\omega+\omega'\right]s}\Psi^{*}(s+2M_0\Delta_{\omega}(x))\Psi(s-2M_0\Delta_{\omega}(x)),
\]
with $\Delta_{\omega}(x)\equiv x-\bar{x}_\omega(x)$. To better connect this expression with the classical case we can make the general assumption that the wave-packet $\Psi$ of the shell is centered in time $\bar{v}_0$ and mass $\bar{M}$. We define $\Phi$ such that

\begin{equation}
\Psi(v_0)\equiv\Phi(v_0-\bar{v}_0)e^{-i\bar{M}\frac{v_0-\bar{v}_0}{\hbar}}\label{eq:redef_wavefunction_shell}.
\end{equation}
 Now 
\[
\left\langle \hat{\beta}\right\rangle_{\omega\omega'}=-\frac{2M_{0}e^{-i\left[\omega+\omega'\right]\bar{v}_0}}{\pi}\sqrt{\frac{\omega'}{\omega}}{\int_0^\infty}dx\sqrt{\frac{\left|\ln(\bar{x}_\omega\left(x\right))\right|}{\left|\ln(x)\right|}}e^{i4M_0\left[\omega+\omega'\right]x}e^{-i2M_0\left[\omega+\omega'\right]\Delta_{\omega}(x)}e^{i\frac{4\bar{M}M_0}{\hbar}\Delta_{\omega}(x)}\times
\]
\begin{equation}
\times{\int_{-\infty}^\infty}ds e^{-i\left[\omega+\omega'\right]s}\Phi^{*}(s+2M_0\Delta_{\omega}(x))\Phi(s-2M_0\Delta_{\omega}(x))\label{eq:coef_bogol_cuanticos_final}.
\end{equation}

As a possible wave-function for the black hole we consider a Gaussian centered in $\bar{v}_0$ and $\bar{M}_0$ whose $v_0$ representation is

\begin{equation}
\Psi_{b}\left(v_{0}\right)=\frac{1}{\left(\pi\sigma^{2}\right)^{\frac{1}{4}}}e^{-\frac{\left(v_{0}-\bar{v}_{0}\right)^{2}}{2\sigma^{2}}}e^{-i\bar{M}\frac{v_{0}-\bar{v_{0}}}{\hbar}}\label{eq:gaussiana}.
\end{equation}

Using this wave-function we get 
\[
\left\langle \hat{\beta}\right\rangle_{\omega\omega'}=-\frac{2M_{0}e^{-i\left[\omega+\omega'\right]\bar{v}_0}}{\pi}\sqrt{\frac{\omega'}{\omega}}{\int_0^\infty}dx\sqrt{\frac{\left|\ln(\bar{x}_\omega\left(x\right))\right|}{\left|\ln(x)\right|}}e^{i4M_0\left[\omega+\omega'\right]x}e^{-i2M_0\left[\omega+\omega'\right]\Delta_{\omega}(x)}e^{i\frac{4\bar{M}M_{0}}{\hbar}\Delta_{\omega}(x)}\times
\]
\[
\times\frac{1}{\left(\pi\sigma^{2}\right)^{\frac{1}{2}}}e^{-\frac{4M_0^2\Delta_{\omega}(x)^2}{\sigma^2}} {\int_{-\infty}^\infty}ds e^{-i\left[\omega+\omega'\right]s}e^{-\frac{s^2}{\sigma^2}}.
\]

Computing the Gaussian integral
\[
\left\langle \hat{\beta}\right\rangle_{\omega\omega'}=-\frac{2M_{0}}{\pi}\sqrt{\frac{\omega'}{\omega}}e^{-i\left[\omega+\omega'\right]\bar{v}_{0}}e^{-\left[\omega+\omega'\right]^{2}\frac{\sigma^{2}}{4}}{\int_0^\infty}dx\sqrt{\frac{\left|\ln(\bar{x}_\omega\left(x\right))\right|}{\left|\ln(x)\right|}}\times
\]
\begin{equation}
\times e^{i4M_{0}\left[\omega+\omega'\right]x} e^{i\frac{4\bar{M}M_{0}}{\hbar}\Delta_{\omega}(x)}e^{-\frac{4M_{0}^{2}}{\sigma^{2}}\Delta_{\omega}(x)^{2}}e^{-i2M_{0}\left[\omega+\omega'\right]\Delta_{\omega}(x)}\label{eq:coef_bogol_gaussiana}.
\end{equation}

To check the consistency of this result we can get the classical limit
by taking $\hbar$ to zero and the width of the packet in both
canonical variables to zero as well,
\begin{equation}
\hbar \to 0, \ \sigma \to 0 \quad   \text{with } \quad \frac{\hbar}{\sigma} \to 0 . \label{classlim}
\end{equation}
In that limit 
$\bar{x}_\omega(x)={\rm li}^{-1}\left[{\rm li}(x)-\frac{\omega\hbar}{M_{0}}\right]\to
x$ and 
$\frac{\Delta_{\omega}(x)}{\hbar}\to\frac{\omega}{M_{0}}\ln(x)$. Therefore,
\[
\left\langle \hat{\beta}\right\rangle_{\omega\omega'}\underset{\hbar\to0}{\longrightarrow}-\frac{4M_{0}}{2\pi}\sqrt{\frac{\omega'}{\omega}}e^{-i\left[\omega+\omega'\right]\bar{v_{0}}}{\int_0^\infty}dxe^{4M_{0}i\left[\omega+\omega'\right]x}e^{i4\bar{M}\omega \ln(x)}=\beta_{\omega\omega'}
\]
and we recover the classical expression (\ref{betaclasico}).

\subsection{Corrections to Hawking radiation: a first approach}

In the previous subsection we obtained the Bogoliubov coefficients in
the full quantum treatment and showed that we recover the classical
result in the classical limit (\ref{classlim}). Here we would like to
study deviations from the classical behaviour. For it, we will use the
expectation values derived in the previous section. This is only a
first approximation since the correct expression involves the
expectation value of products of the operators associated with the
Bogoliubov coefficients. We will later see that this implies an
important difference and an interesting example of how the quantum
fluctuations may be determinant and lead to significant departures
from the mean field approach.

We will consider the
example of a Gaussian wave-packet for the wave-function of the shell and
arrive to some general conclusions. Then, to maintain tractable
expressions, we will restrict attention to ``extreme'' cases of the
latter: one with the Gaussian very peaked in mass (with large
dispersion in $v_0$) and the other with the Gaussian very peaked in
$v_0$ (with large dispersion in the mass).

Let us start with some general considerations about the expectation
value of the operator associated with the Bogoliubov coefficients.
Expression (\ref{eq:coef_bogol_gaussiana}) has several differences with the classical limit (\ref{betaclasico}), especially in the dependence with the frequency $\omega'$. Lets focus in the integrand 

$$\sqrt{\frac{\left|\ln(\bar{x}_\omega\left(x\right))\right|}{\left|\ln(x)\right|}}e^{i4M_{0}\left[\omega+\omega'\right]x} e^{i\frac{4\bar{M}M_{0}}{\hbar}\Delta_{\omega}(x)}e^{-\frac{4M_{0}^{2}}{\sigma^{2}}\Delta_{\omega}(x)^{2}}e^{-i2M_{0}\left[\omega+\omega'\right]\Delta_{\omega}(x)}.$$
Taking into account that
$$\Delta_{\omega}(x)\underset{x\to0}{\longrightarrow} {\rm li}^{-1}\left(-\frac{\hbar\omega}{M_0}\right)$$
 $$\Delta_{\omega}(x)\underset{x\to+\infty}{\sim} \frac{\hbar\omega}{M_0}\ln(x),$$
 and remembering that
 $$\bar{x}_\omega(x)={\rm li}^{-1}\left({\rm li}(x)-\frac{\hbar\omega}{M_0}\right),$$
we see it vanishes when $x\to0$, also $\sqrt{\frac{\left|\ln(\bar{x}_\omega\left(x\right))\right|}{\left|\ln(x)\right|}}$ is bounded by $1$ and finally
 $$e^{-\frac{4M_{0}^{2}}{\sigma^{2}}\Delta_{\omega}(x)^{2}}\underset{x\to+\infty}{\sim}e^{-\frac{4\hbar^2}{\sigma^{2}}\ln(x)^{2}}.$$
 
 Therefore the integral has a bound (independent of $\omega'$) given by
 
 $${\int_0^\infty}dxe^{-\frac{4M_{0}^{2}}{\sigma^{2}}\Delta_{\omega}(x)^{2}}.$$ 
 This fact, together with the exponential factor
 $e^{-\left[\omega+\omega'\right]^{2}\frac{\sigma^{2}}{4}}$ outside
 the integral, ensures exponential suppression of large $\omega'$
 contributions. The integral also lacks the $\frac{1}{\omega'+\omega}$
 dependence that the classical expression has since setting
 $\omega'=\omega=0$ inside the integral still gives us a finite result.

 One quantity that is extremely sensitive to these differences is
 the total number of emitted particles per unit frequency. If we compute it using the
 expectation value of the Bogoliubov coefficients it will be given
 by
\begin{equation}
\left\langle N_{\omega}^{AQS}\right\rangle ={\int_0^\infty}d\omega'\left\langle \hat{\beta}\right\rangle _{\omega\omega'}\left\langle \hat{\beta}\right\rangle _{\omega\omega'}^{*}\label{eq:N_AQS}
\end{equation}
where the superscript ``AQS'' stands for Approximate Quantum
Shell. The reason to call it approximate is that the correct way to
compute it would be with the expectation value of the product of
Bogoliubov coefficients instead of the product of expectation
values. We will address this important issue in the next section,
but for now we will assume that fluctuations are small and this is a good
approximation.

Given the previous general remarks about Bogoliubov coefficients we
conclude $\left\langle N_{\omega}^{AQS}\right\rangle$ is not divergent
as in the classical expression (\ref{eq:N_omega3}) but finite which is
a big departure from eternal Hawking radiation.\\
 
A more explicit analysis can be performed with a state that is
squeezed with large dispersion in the position of the shell and very
peaked in the mass. 
Specifically, we will consider the case where the shell is in a Gaussian (\ref{eq:gaussiana}) squeezed state with large dispersion in $v_0$ and small dispersion in $M$. The leading quantum correction for such states is obtained by taking the limit $\hbar \to 0$ with
\begin{equation}
 \Delta v_0 = \sigma =  \text{constant} = Z \ell_{\text{Planck}} \; , \; Z \gg 1 \, ; \quad \Delta M= \hbar/\sigma.\label{eq:squeezed_state_limit}
\end{equation}
Even though this limit is
different from the one we took following (\ref{eq:coef_bogol_gaussiana})  it has similarities with it. The
terms inside the integral go to their classical values but the
external factor involving $\sigma$ now remains. One then finds that (\ref{eq:coef_bogol_gaussiana})  goes to:
\begin{equation}
\left\langle \hat{\beta}\right\rangle_{\omega\omega'}\to e^{-\left[\omega+\omega'\right]^{2}\frac{\sigma^{2}}{4}}\beta_{\omega\omega'}. \label{betasq}
\end{equation}
The deviation from the classical Bogoliubov coefficients is only
through a multiplicative factor that disappears in the classical limit
where $\sigma \to 0$. For non-zero $\sigma$ the factor suppresses
frequencies greater than $1/\sigma$. This  produces important
corrections to the calculation of Hawking radiation as we already
mentioned. However this calculation is based in an approximation in
which we computed the square of the expectation value of the
Bogoliubov coefficients instead of the expectation value of the
square. It turns out this approximation breaks down. We present
detailed calculations in the appendix. Here we just outline the
calculation.

Estimating the expectation value of the number operator using expression (\ref{betasq}) we get,
\[
\left\langle N_{\omega}^{AQS}\right\rangle ={\int_0^\infty}d\omega'\left\langle \hat{\beta}\right\rangle _{\omega\omega'}\left\langle \hat{\beta}\right\rangle _{\omega\omega'}^{*}=\frac{1}{4\pi^{2}}\frac{1}{\omega}\left|\Gamma(1+4\bar{M}\omega i)\right|^{2}e^{-4\pi \bar{M}\omega}{\int_0^\infty}d\omega'\frac{\omega'e^{-\left[\omega+\omega'\right]^{2}\frac{\sigma^{2}}{2}}}{\left(\omega'+\omega\right)^{2}}=\]

\begin{equation}
=\frac{1}{e^{8\bar{M}\pi\omega}-1}\frac{2\bar{M}}{\pi}{\int_0^\infty}d\omega'\frac{\omega'e^{-\left[\omega+\omega'\right]^{2}\frac{\sigma^{2}}{2}}}{\left(\omega'+\omega\right)^{2}}.\label{eq:Number of particles_AQS}
\end{equation}
This expression has the same pre-factor Hawking radiation has but with $\bar{M}$ in the role of mass. However, unlike (\ref{eq:N_omega3}) this is a finite expression for all $\omega\neq0$ and has a logarithmic divergence when $\omega\to0$. Furthermore, it has a $\exp(-\frac{\omega^2\sigma^2}{2})$ dependence when $\omega\to+\infty$ instead of the usual $\exp(-8\bar{M}\pi\omega)$ for Hawking radiation. \\

\begin{figure}
\includegraphics[height=8cm]{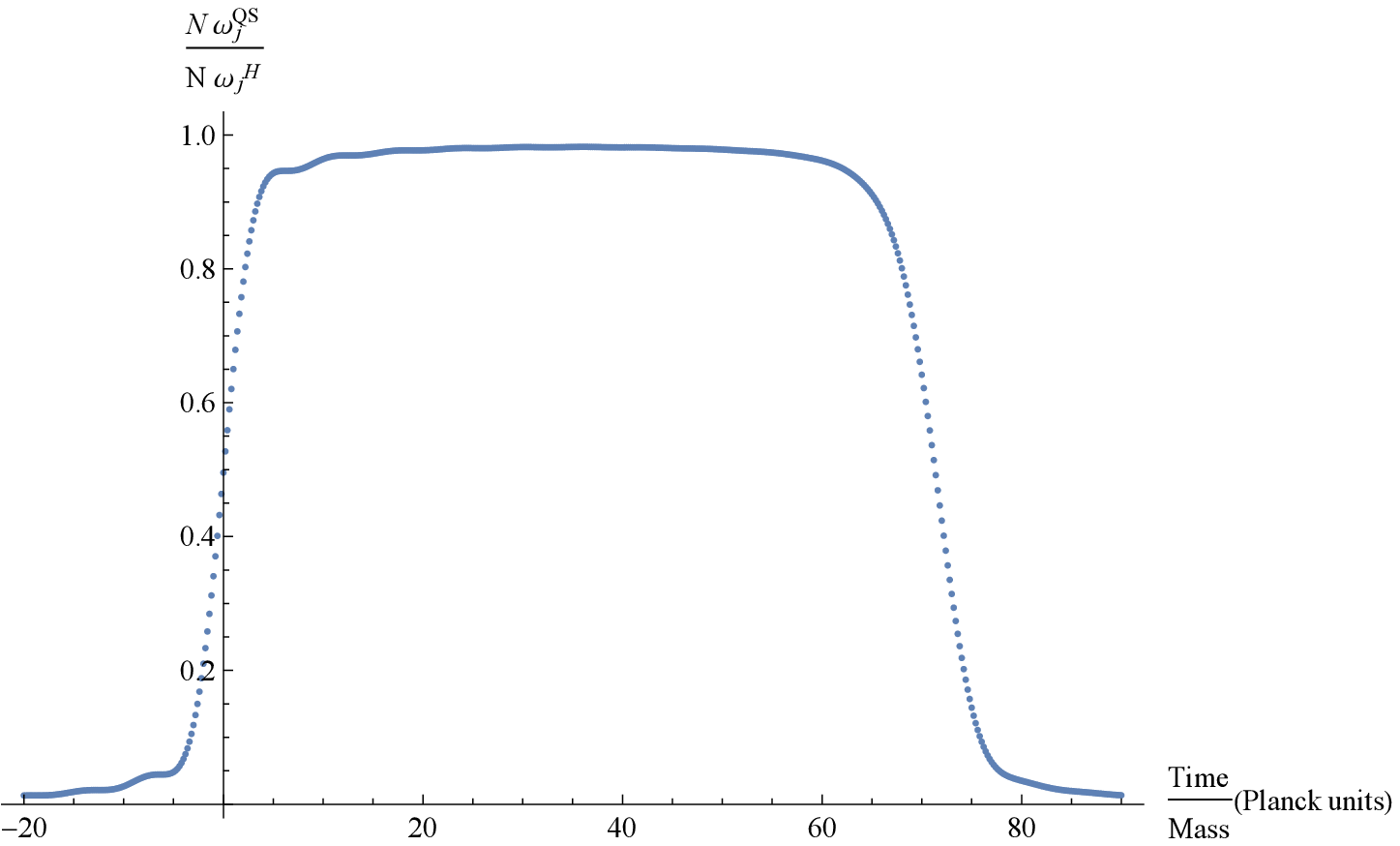}
\caption{This plot shows the departure from the classical result of $N^{AQS}_{\omega_j}/N^{H}_{\omega_j}$. We have considered $\omega_j$ corresponding to the $\lambda$ of maximum emission ($\lambda_m \sim 16 R_s$), the frequency interval $\epsilon = c/R_s$ and the shell's position uncertainty $\sigma= 5 R_s \times 10^{-38}$ ($\sim 3 l_P$ for $R_s= 1 km$). Note that the time step is $2\pi/\epsilon$.}
\label{nwj}
\end{figure}

Since we are interested in the behaviour of the Hawking radiation as a
function of time it is convenient to introduce wave packets as we
considered before and therefore to compute the number of particles at
time $u_n$ around $\omega_j$ given by,
\[
\langle N_{\omega_{j}}^{AQS}\rangle=\frac{1}{\epsilon}
{\int_{j\epsilon}^{\left(j+1\right)\epsilon}\int_{j\epsilon}^{\left(j+1\right)\epsilon}}d\omega_{1}d\omega_{2}e^{-u_n\Delta\omega i}\left\langle\rho^{AQS}_{\omega_1,\omega_2}\right\rangle.
\]
Using the results in appendix 2 it can be computed explicitly,
yielding, 
\[
\langle N_{\omega_{j}}^{AQS}\rangle
=\frac{\bar{M}\epsilon}{\pi}\frac{1}{e^{8\bar{M}\pi\omega_{j}}-1}
{\int_1^\infty}dy\frac{e^{-\frac{\omega_{j}^{2}\sigma^{2}}{2}y}}{y}\left\{\frac{\sin\left[\frac{\epsilon}{2}\left(\alpha-2\bar{M}\ln\left(y\right)\right)\right]}{\frac{\epsilon}{2}\left(\alpha-2\bar{M}\ln\left(y\right)\right)}\right\} ^{2}.
\]
Where $\alpha$ is the same quantity defined in equation (\ref{alpha})
with $M$ and $v_0$ replaced by their respective expectation values in
the Gaussian state given above.
 The presence of the factor 
 $\sin^{2}(a)/a^{2}$ 
and the decreasing exponential imply that the integral decreases when 
$\alpha$ grows and also drastically decreases when 
$\alpha<0$. The latter is a result we already knew from the classical
case, but the former is a result of the quantum nature of the black
hole since it is not present if 
$\sigma=0$. Figure (\ref{nwj}) shows the departure from the classical result that appears when one computes the frequency distribution starting from $\left\langle\hat{\beta}\right\rangle$.
We can estimate the time of emission for each frequency using both
extremes. In the appendix we also show that the features are robust
with respect to the choice of the quantum state by considering
squeezed
states with large dispersion in the mass, which is the opposite of the
choice we considered here. 
However, as we shall see in the next section, the decrease
in emission for late time is an artifact of the approximation
considered that neglects the fluctuations of the number of particles. 

\section{Computing the expectation value of the density matrix in the
  complete quantum treatment} 

In this section we will obtain an exact expression for the expectation
value of the density matrix with the same technique used to compute
the expectation value of Bogoliubov coefficients. From its diagonal
terms we can compute the number of particles produced as a function of
frequency.

From expression ({\ref{eq:coef_bogol_cuanticos}) for the operator associated to a Bogoliubov coefficient we can compute the expectation value of the density matrix as

\[
\left\langle  \rho_{\omega_1\omega_2}^{QS}\right\rangle=\int_{0}^{\infty} d\omega'\left\langle\hat{\beta}_{\omega_1\omega'}\hat{\beta}_{\omega_2\omega'}^{*}\right\rangle.
\]
where $QS$ stands for quantum shell. The full expression is
\[
\left\langle  \rho_{\omega_1\omega_2}^{QS}\right\rangle=\frac{1}{(2\pi)^2}\int_{0}^{\infty} d\omega'\frac{\omega'}{\sqrt{\omega_1\omega_2}}\left\langle
  \Psi\right|
{{\int\int}_{-\infty}^{+\infty}}dvdv'{\int_{-\infty}^{+\infty}}dv_{0}\left|v_{0}\right\rangle \left\langle v_{0}\right|\theta\left(\widehat{v}_{0}-v\widehat{I}\right)e^{-i\omega_1\hat{u}(v)- i\omega'v}\times
\]
\[
\times \underset{J=1,2}{\sum}{\int_{-\infty}^{+\infty}}du\left|u,J\right\rangle\left\langle u,J\right|{\int_{-\infty}^{+\infty}}dv''_{0}\left|v''_{0}\right\rangle \left\langle v''_{0}\right|\theta\left(\widehat{v}_{0}-v\widehat{I}\right){\int_{-\infty}^{+\infty}}dv'''_{0}\left|v'''_{0}\right\rangle \left\langle v'''_{0}\right|\theta\left(\widehat{v}_{0}-v'\widehat{I}\right)\times
\]
\[
\times\underset{L=1,2}{\sum}{\int_{-\infty}^{+\infty}}du'\left|u',L\right\rangle\left\langle u',L\right| e^{i\omega_2\hat{u'}(v')+ i\omega'v'}{\int_{-\infty}^{+\infty}}dv'_{0}\left|v'_{0}\right\rangle \left\langle v'_{0}\right|\theta\left(\widehat{v}_{0}-v'\widehat{I}\right)\left|\Psi\right\rangle.
\]
 Here we have considered bases of eigenstates of $\hat{v}_{0}$ and $\hat{u}$ and we have omitted the $\epsilon$ dependence in $\hat{u}$ eigenstates. Identical arguments as the ones used in the appendix allow us to do so. Simplifying the expression we get
\[
\left\langle  \rho_{\omega_1\omega_2}^{QS}\right\rangle=\frac{1}{(2\pi)^2}\int_{0}^{\infty} d\omega'\frac{\omega'}{\sqrt{\omega_1\omega_2}}
{{\int}_{-\infty}^{+\infty}}dvdv'dv_0dv'_0dv''_0dv'''_0\theta\left(v_{0}-v\right)\theta\left(v''_{0}-v\right)\theta\left(v'''_{0}-v'\right)\theta\left(v'_{0}-v'\right)\delta\left(v''_{0}-v'''_0\right)\times
\]
\[
\times{\int}_{-\infty}^{+\infty}dudu' e^{-i\omega_1 u - i\omega'v}\underset{J=1,2}{\sum}\psi_{u,J}(v_{0})\psi_{u,J}^{*}(v''_{0}) e^{i\omega_2 u' + i\omega'v'}\underset{L=1,2}{\sum}\psi_{u',L}(v'''_{0})\psi_{u',L}^{*}(v'_{0})\Psi^{*}(v_0)\Psi(v'_0), 
\]

The change of variables $x_{1}=\frac{v_{0}-v}{4M_{0}}$, $x_{2}=\frac{v''_{0}-v}{4M_{0}}$, $x_{3}=\frac{v'''_{0}-v'}{4M_{0}}$ y $x_{4}=\frac{v'_{0}-v'}{4M_{0}}$
take us to

\[
\left\langle  \rho_{\omega_1\omega_2}^{QS}\right\rangle=\frac{\left(4M_0\right)^4}{(2\pi)^2}\int_{0}^{\infty} d\omega'\frac{\omega'}{\sqrt{\omega_1\omega_2}}
{{\int}_{-\infty}^{+\infty}}dvdv'{\int}_{0}^{+\infty}dx_1dx_2dx_3dx_4\delta\left(4M_0\left[x_2-x_3\right]+v-v'\right)\times
\]
\[
\times{\int}_{-\infty}^{+\infty}dudu' e^{-i\omega_1 u - i\omega'v}\underset{J=1,2}{\sum}\psi_{u,J}(x_1)\psi_{u,J}^{*}(x_2) e^{i\omega_2 u' + i\omega'v'}\underset{L=1,2}{\sum}\psi_{u',L}(x_3)\psi_{u',L}^{*}(x_4)\Psi^{*}(4M_{0}x_{1}+v)\Psi(4M_{0}x_{4}+v'). 
\]

Using expressions (\ref{eq:autoestados_de_u_1}) and (\ref{eq:autoestados_de_u_2}) for the eigenfunctions of the $\hat{u}$ operator

\[
\left\langle  \rho_{\omega_1\omega_2}^{QS}\right\rangle=\frac{\left(4M_0\right)^4}{(16\pi^2\hbar)^2}\int_{0}^{\infty} d\omega'\frac{\omega'}{\sqrt{\omega_1\omega_2}}
{{\int}_{-\infty}^{+\infty}}dvdv'{\int}_{0}^{+\infty}dx_1dx_2{\int}_{0}^{+\infty}dx_3dx_4\delta\left(4M_0\left[x_2-x_3\right]+v-v'\right){\int}_{-\infty}^{+\infty}dudu' \times
\]
\[
\times e^{-i\omega_1 u - i\omega'v}e^{i\omega_2 u' + i\omega'v'}\frac{\exp\left(\frac{iM_{0}}{\hbar}(u-v)\left[{\rm
      li}(x_{1})-{\rm li}(x_{2})\right]\right)\exp\left(\frac{iM_{0}}{\hbar}(u'-v')\left[{\rm
      li}(x_{3})-{\rm li}(x_{4})\right]\right)}{\sqrt{\left|\ln(x_{1})\right|\left|\ln(x_{2})\right|\left|\ln(x_{3})\right|\left|\ln(x_{4})\right|}}\Psi^{*}(4M_{0}x_{1}+v)\Psi(4M_{0}x_{4}+v'). 
\]

Integrating in $u$ y $u'$ we get,

\[
\left\langle  \rho_{\omega_1\omega_2}^{QS}\right\rangle=\frac{\left(4M_0\right)^4}{(8\pi\hbar)^2}\int_{0}^{\infty} d\omega'\frac{\omega'}{\sqrt{\omega_1\omega_2}}
{{\int}_{-\infty}^{+\infty}}dvdv'{\int}_{0}^{+\infty}dx_1dx_2{\int}_{0}^{+\infty}dx_3dx_4\delta\left(4M_0\left[x_2-x_3\right]+v-v'\right)\times
\]
\[
\times e^{- i\left[\omega'+\omega_1\right]v}e^{i\left[\omega'+\omega_2\right]v'}\frac{\delta\left(\omega_1-\frac{M_{0}}{\hbar}\left[{\rm
      li}(x_{1})-{\rm li}(x_{2})\right]\right)\delta\left(\omega_2+\frac{M_{0}}{\hbar}\left[{\rm
      li}(x_{3})-{\rm li}(x_{4})\right]\right)}{\sqrt{\left|\ln(x_{1})\right|\left|\ln(x_{2})\right|\left|\ln(x_{3})\right|\left|\ln(x_{4})\right|}}\Psi^{*}(4M_{0}x_{1}+v)\Psi(4M_{0}x_{4}+v'). 
\]

Since ${\rm li}$ is invertible in $\left(0,1\right)$ and in $\left(1,+\infty\right)$
we can integrate in $x_{2}$ and $x_3$ to get 

\[
\left\langle  \rho_{\omega_1\omega_2}^{QS}\right\rangle=\frac{\left(2M_0\right)^2}{\pi^2}\int_{0}^{\infty} d\omega'\frac{\omega'}{\sqrt{\omega_1\omega_2}}
{{\int}_{-\infty}^{+\infty}}dvdv'{\int}_{0}^{+\infty}dx_1{\int}_{0}^{+\infty}dx_4\delta\left(4M_0\left[x_2(x_1)-x_3(x_4)\right]+v-v'\right)\times
\]
\[
\times e^{- i\left[\omega'+\omega_1\right]v}e^{i\left[\omega'+\omega_2\right]v'}\sqrt{\frac{\left|\ln(x_{2}(x_1)\right|\left|\ln(x_{3}(x_4))\right|}{\left|\ln(x_{1})\right|\left|\ln(x_{4})\right|}}\Psi^{*}(4M_{0}x_{1}+v)\Psi(4M_{0}x_{4}+v'),
\]
where $x_{2}\left(x_{1}\right)={\rm li}^{-1}\left[{\rm li}\left(x_{1}\right)-\frac{\omega_1\hbar}{M_{0}}\right]$, $x_{3}\left(x_{4}\right)={\rm li}^{-1}\left[{\rm li}\left(x_{4}\right)-\frac{\omega_2\hbar}{M_{0}}\right]$
and we have used that  $\partial_t{\rm li}\left(t\right)=\frac{1}{\left|\ln(t)\right|}$.
We redefine $x=x_{1}$, $x'=x_{4}$  and then

\[
\left\langle  \rho_{\omega_1\omega_2}^{QS}\right\rangle=\frac{\left(2M_0\right)^2}{\pi^2}\int_{0}^{\infty} d\omega'\frac{\omega'}{\sqrt{\omega_1\omega_2}}
{{\int}_{-\infty}^{+\infty}}dvdv'{\int}_{0}^{+\infty}dxdx'\delta\left(4M_0\left[\bar{x}_{\omega_1}(x)-\bar{x}_{\omega_2}(x')\right]+v-v'\right)\times
\]
\[
\times e^{- i\left[\omega'+\omega_1\right]v}e^{i\left[\omega'+\omega_2\right]v'}\sqrt{\frac{\left|\ln(\bar{x}_{\omega_1}(x))\right|\left|\ln(\bar{x}_{\omega_2}(x'))\right|}{\left|\ln(x)\right|\left|\ln(x')\right|}}\Psi^{*}(4M_{0}x+v)\Psi(4M_{0}x'+v')
.\]

Integrating in $v'$

\[
\left\langle  \rho_{\omega_1\omega_2}^{QS}\right\rangle=\frac{\left(2M_0\right)^2}{\pi^2}\int_{0}^{\infty} d\omega'\frac{\omega'}{\sqrt{\omega_1\omega_2}}
{\int}_{0}^{+\infty}dxdx'e^{-i4M_0\left[\omega'+\omega_2\right]x'}e^{i4M_0\left[\omega'+\omega_2\right]x}
e^{i4M_0\left[\omega'+\omega_2\right]\Delta_{\omega_1\omega_2}(x,x')}\times
\]
\[
\times\sqrt{\frac{\left|\ln(\bar{x}_{\omega_1}(x))\right|\left|\ln(\bar{x}_{\omega_2}(x'))\right|}{\left|\ln(x)\right|\left|\ln(x')\right|}} {{\int}_{-\infty}^{+\infty}}dv e^{- i\left[\omega_1-\omega_2\right]v}\Psi^{*}(4M_{0}x+v)\Psi(4M_{0}x+v+4M_0\Delta_{\omega_1\omega_2}(x,x'))
\]

where $\Delta_{\omega_1\omega_2}(x,x')=\Delta_{\omega_2}(x')-\Delta_{\omega_1}(x)$. Now, changing variable $v$ to $s=v+4M_0x+2M_0\Delta_{\omega_1\omega_2}(x,x')$

\[
\left\langle  \rho_{\omega_1\omega_2}^{QS}\right\rangle=\frac{\left(2M_0\right)^2}{\pi^2}\int_{0}^{\infty} d\omega'\frac{\omega'}{\sqrt{\omega_1\omega_2}}
{\int}_{0}^{+\infty}dxdx'e^{-i4M_0\left[\omega'+\omega_2\right]x'}e^{i4M_0\left[\omega'+\omega_1\right]x}
e^{i4M_0\left[\omega'+\bar{\omega}\right]\Delta_{\omega_1\omega_2}(x,x')}\times
\]
\[
\times\sqrt{\frac{\left|\ln(\bar{x}_{\omega_1}(x))\right|\left|\ln(\bar{x}_{\omega_2}(x'))\right|}{\left|\ln(x)\right|\left|\ln(x')\right|}} {{\int}_{-\infty}^{+\infty}}ds e^{- i\left[\omega_1-\omega_2\right]s}\Psi^{*}(s-2M_0\Delta_{\omega_1\omega_2}(x,x'))\Psi(s+2M_0\Delta_{\omega_1\omega_2}(x,x'))
\]

where $\bar{\omega}=\frac{\omega_1+\omega_2}{2}$. Finally, using definition (\ref{eq:redef_wavefunction_shell}) we get

\[
\left\langle  \rho_{\omega_1\omega_2}^{QS}\right\rangle=\frac{\left(2M_0\right)^2e^{- i\left[\omega_1-\omega_2\right]\bar{v}_0}}{\pi^2\sqrt{\omega_1\omega_2}}\int_{0}^{\infty} d\omega'\omega'
{\int}_{0}^{+\infty}dxdx'e^{-i4M_0\left[\omega'+\omega_2\right]x'}e^{i4M_0\left[\omega'+\omega_1\right]x}
e^{i4M_0\left[\omega'+\bar{\omega}\right]\Delta_{\omega_1\omega_2}(x,x')} \times
\]
\begin{equation}
\times e^{-i\frac{4M_0\bar{M}}{\hbar}\Delta_{\omega_1\omega_2}(x,x')}\sqrt{\frac{\left|\ln(\bar{x}_{\omega_1}(x))\right|\left|\ln(\bar{x}_{\omega_2}(x'))\right|}{\left|\ln(x)\right|\left|\ln(x')\right|}} {{\int}_{-\infty}^{+\infty}}ds e^{- i\left[\omega_1-\omega_2\right]s}\Phi^*(s-2M_0\Delta_{\omega_1\omega_2}(x,x'))
\Phi(s+2M_0\Delta_{\omega_1\omega_2}(x,x')).
\label{eq:matriz_densidad_quantum}
\end{equation}

where $\Delta\omega=\omega_2-\omega_1$. Taking again the Gaussian wavepacket (\ref{eq:gaussiana}) as an example, we get
\[
\left\langle  \rho_{\omega_1\omega_2}^{QS}\right\rangle=\frac{\left(2M_0\right)^2e^{ i\Delta\omega\bar{v}_0}e^{-\frac{\Delta\omega^2\sigma^2}{4}}}{\pi^2\sqrt{\omega_1\omega_2}}\int_{0}^{\infty} d\omega'\omega'
{\int}_{0}^{+\infty}dxdx'e^{i4M_0\left[\omega'+\omega_1\right]x}e^{-i4M_0\left[\omega'+\omega_2\right]x'}\times
\]
\begin{equation}
\times e^{i4M_0\left[\omega'+\bar{\omega}\right]\Delta_{\omega_1\omega_2}(x,x')} e^{-i\frac{4M_0\bar{M}}{\hbar}\Delta_{\omega_1\omega_2}(x,x')}\sqrt{\frac{\left|\ln(\bar{x}_{\omega_1}(x))\right|\left|\ln(\bar{x}_{\omega_2}(x'))\right|}{\left|\ln(x)\right|\left|\ln(x')\right|}}e^{-\frac{4M_0^2\Delta_{\omega_1\omega_2}(x,x')^2}{\sigma^2}}.\label{eq:matriz_densidad_quantum_gaussiana}
\end{equation}
This is the final result for the expectation value of the density
matrix in the complete quantum treatment. 

From this expression we can compute the classical limit (\ref{classlim}). 
In that limit 
$\bar{x}_{\omega_1}(x)\to x$, $\bar{x}_{\omega_2}(x')\to x'$ and $\frac{\Delta_{\omega_1\omega_2}(x,x')}{\hbar}\to\frac{\omega_2\ln(x')-\omega_1\ln(x)}{M_{0}}$. Therefore,
\[
\left\langle  \rho_{\omega_1\omega_2}^{QS}\right\rangle=\frac{\left(2M_0\right)^2e^{ i\Delta\omega\bar{v}_0}}{\pi^2\sqrt{\omega_1\omega_2}}\int_{0}^{\infty} d\omega'\omega'
{\int}_{0}^{+\infty}dxe^{i4M_0\left[\omega'+\omega_1\right]x}e^{i4\bar{M}\omega_1\ln(x)}{\int}_{0}^{+\infty}dx'e^{-i4M_0\left[\omega'+\omega_2\right]x'}e^{-i4\bar{M}\omega_2\ln(x')}
\]
which is the classical expression for the density matrix
\[
\rho^{CS}_{\omega_1,\omega_2}={\int_0^\infty}d\omega'\beta_{\omega_{1}\omega'}\beta_{\omega_{2}\omega'}^{*}\]
with $\beta_{\omega\omega'}$ given by (\ref{betaclasico}).

We analyze the consequences of these calculations in the next section.

\section{Corrections to Hawking radiation due to the quantum
  background} 

We have studied the corrections to Hawking radiation using the
approximate expression (\ref{eq:approx_density matrix}) discussed in
appendix 2. Now we can do
the same calculation from the exact expression
(\ref{eq:matriz_densidad_quantum_gaussiana}). As in the previous
section we begin with some general remarks about the result for a
Gaussian state and then explore the same squeezed states we considered
before.

Unlike the density matrix constructed from
(\ref{eq:coef_bogol_gaussiana}), expression
(\ref{eq:matriz_densidad_quantum_gaussiana}) has a double integral
that can not be separated in $x$ and $x'$ variables. But the most
significant differences are the missing $\omega'$ dependence in the
exponential
$$e^{-\frac{\Delta\omega^2\sigma^2}{4}},$$ and the exponential inside
the 
double integral 
$$e^{-\frac{4M_0^2\left[\Delta_{\omega_1\omega_2}(x,x')\right]^2}{\sigma^2}}.$$
The first point significantly changes the $\omega'$ integral. The
second expression does not make the integrand fall rapidly when
$x,x'\to+\infty$ because the exponential remains constant in the
directions give by the equation
$$\Delta_{\omega_1\omega_2}(x,x')=\Delta_{\omega_2}(x')-\Delta_{\omega_1}(x)={\rm
  const.}$$

As we will see in better detail with the following examples, the
consequence of the above remarks are that radiation does not end at a
finite time as predicted by evaluations of the expectation value of
Bogoliubov coefficients. However, the significant difference between
$\left\langle N^{QS}_\omega\right\rangle$ and $\left\langle
  N^{AQS}_\omega\right\rangle$ is also generically associated with  
the appearance of fluctuations in the Bogoliubov coefficients at
finite time. We will see that may leads to new correlations in the
Hawking radiation that are not present in the classical calculation.

\subsection{States peaked in the mass recover the classical results}

Let us consider first the case of a squeezed state with large
dispersion in the position of the shell. 
Taking the limit (\ref{eq:squeezed_state_limit}),  
\begin{equation}
\left\langle  \rho_{\omega_1\omega_2}^{QS}\right\rangle\to e^{-\frac{\Delta\omega^2\sigma^2}{4}}\left\langle  \rho_{\omega_1\omega_2}^{CS}\right\rangle.
\end{equation}

It is clear that there are no corrections to the total number of
particles $\left\langle
  N^{QS}_\omega\right\rangle=\left\langle\rho_{\omega\omega}^{QS}\right\rangle$
since the exponential factor is one if $\omega_1=\omega_2$. 
Also, for late times $\left\langle
  \rho_{\omega_1\omega_2}^{CS}\right\rangle$ is diagonal so the are no
non-vanishing correlations for different frequencies. We therefore
recover the classical results in their entirety for the particular
case of squeezed states we consider that are highly peaked in the mass
and with large dispersion in the position of the shell.

\subsection{States with dispersion in the mass}

To illustrate this point 
let us consider now a squeezed state with large dispersion in the mass
of the shell.  To compare with the previous result let us compute the
number of particles taking the limit
(\ref{eq:squeezed_state_M_limit}). We get,
\begin{equation}
\left\langle N_{\omega}^{QS}\right\rangle=\left\langle\rho_{\omega\omega}^{QS}\right\rangle\to\frac{\left(2M_0\right)^2}{\pi^2\omega}\int_{0}^{\infty} d\omega'\omega'
{\int}_{0}^{+\infty}dxdx'e^{-\epsilon\left(x+x'\right)}e^{-i4M_0\left[\omega'+\omega\right]\left(x'-x\right)}e^{-i4\bar{M}\omega\ln\left(\frac{x'}{x}\right)}e^{-4\Delta M^2\omega^2\ln\left(\frac{x'}{x}\right)^2}\label{eq:matriz_densidad_quantum_gaussiana_dM}
\end{equation}
where we introduced the same $\epsilon$ regulator used for the integration of Bogoliubov coefficients. The change of variables $x=r\cos(\theta)$, $x'=r\sin(\theta)$ allow us to compute the double integral as
\[
\int_0^{\pi/2}d\theta\int_0^{+\infty}r dr e^{-\epsilon r\left[\sin(\theta)+\cos(\theta)\right]}e^{-i4M_0\left[\omega'+\omega\right]r\left[\sin(\theta)-cos(\theta)\right]}e^{-i4\bar{M}\omega\ln\left[\tan(\theta)\right]}e^{-4\Delta M^2\omega^2\ln\left[\tan(\theta)\right]^2}.
\]

The $r$ integral can be computed, leading to,
\[
-\frac{1}{\left[\omega'+\omega\right]^2(4M_0)^2}\underset{\epsilon\to0}{\lim}\int_0^{\pi/2}d\theta\frac{e^{-i4\bar{M}\omega\ln\left[\tan(\theta)\right]}e^{-4\Delta M^2\omega^2\ln\left[\tan(\theta)\right]^2}}{\left[\frac{\tan(\theta)-1}{\tan(\theta)+1}-i\epsilon\right]^2}\frac{1+\tan(\theta)^2}{\left[1+\tan(\theta)\right]^2},
\]
where we have redefined $\epsilon$ conveniently. A final change of variable $y=\ln\left[\tan(\theta)\right]$ turns the integral into

\[
-\frac{1}{\left[\omega'+\omega\right]^2(4M_0)^2}\underset{\epsilon\to0}{\lim}\int_{-\infty}^{+\infty}dy\frac{1}{2\cosh(y/2)}\frac{e^{-i4\bar{M}\omega y}e^{-4\Delta M^2\omega^2y^2}}{\left[\tanh(y/2)-i\epsilon\right]^2}.
\]

This expression can be rewritten as

$$-\frac{1}{4\left[\omega'+\omega\right]^2(4M_0)^2}\left[\underset{\epsilon\to0}{\lim}\int_{-\infty}^{+\infty}dy\frac{1}{\cosh^2(y/2)}\frac{e^{-i4\bar{M}\omega y}}{\left[\tanh(y/2)-i\epsilon\right]^2}-\int_{-\infty}^{+\infty}dye^{-i4\bar{M}\omega y}\frac{1-e^{-4\Delta M^2\omega^2y^2}}{\sinh^2(y/2)}\right].$$

Now the first integral can be computed by contour integration to obtain the classical result (\ref{eq:N_omega3}) with the expectation value $\bar{M}$ in the role of mass,

$$\underset{\epsilon\to0}{\lim}\int_{-\infty}^{+\infty}dy\frac{1}{\cosh^2(y/2)}\frac{e^{-i4\bar{M}\omega y}}{\left[\tanh(y/2)-i\epsilon\right]^2}=\frac{-32\bar{M}\omega\pi}{e^{8\bar{M}\omega\pi}-1}$$

The second term,

$$f(\bar{M},\omega,\Delta M)\equiv \int_{-\infty}^{+\infty}dye^{-i4\bar{M}\omega y}\frac{1-e^{-4\Delta M^2\omega^2y^2}}{\sinh^2(y/2)}$$ 
is a finite correction which vanishes in the classical limit. Regarding the dependence in $\omega$, unlike the leading term it vanishes for $\omega\to0$ and also as the Fourier transform of a smooth and rapidly falling function it falls rapidly with $\omega\to+\infty$. Finally, 

\[
\left\langle N_{\omega}^{QS}\right\rangle=\left[\frac{1}{e^{8\bar{M}\omega\pi}-1}+\frac{f(\bar{M},\omega,\Delta M)}{32\pi\omega\bar{M}}\right]\frac{4\bar{M}}{2\pi}\int_{0}^{\infty} d\omega'\frac{\omega'}{\left[\omega'+\omega\right]^2}.
\]
This expression is clearly divergent, with the same divergent integral
that appears in the classical case but with a small departure from
thermality given by $f$. It could be made finite considering packets
as we did before. Notice that the expression has the thermal spectrum
plus a term that only vanishes when there are no fluctuations in the
mass. The extra term essentially depends on the Fourier transform of
the initial state of the shell and suggests that the complete
information of the initial state could be retrieved from the
radiation.  Recall that in order to recover finite results one needs to
compute the number expectation value for wave packets localized in
time and frequency. We are therefore led to an expression that departs
more and more from ordinary Hawking radiation when the uncertainty in the mass
increases.

\section{Coherence}

Hawking radiation stemming from a classical black hole is
incoherent. This manifests itself in the vanishing of the off-diagonal
elements of the density matrix in the frequency basis. We will see
that the density matrix of the Hawking radiation of the quantum
space-time of the collapsing null shell has non-vanishing off-diagonal
coherence terms which gives additional evidence that it contains
quantum information from the initial state of the shell that gave rise
to the black hole. While
they vanish for standard Hawking radiation on classical space-times 
they are nonvanishing here.

Starting from expression (\ref{eq:matriz_densidad_quantum_gaussiana})
for the density matrix of a Gaussian packet we already discussed the
case of a state extremely peaked in mass and we found no corrections
to the number of particles and no correlations between different
frequencies for late time radiation. On the other hand we studied the
somewhat opposite case of a state  with dispersion in the mass
and well defined position. For that state we found corrections to the
number of particles and now we will study corrections to density
matrix $\rho^{CS}_{\omega_1,\omega_2}$ due to these fluctuations. 
We will only calculate corrections to
the late time density matrix $\rho^{H}_{\omega_1,\omega_2}$. In this
limit the classical matrix is diagonal and therefore the only source
of non diagonal terms will be from the quantum nature of the shell. In
the limit (\ref{eq:squeezed_state_M_limit}) the late time density
matrix takes the form

\[
\left\langle  \rho_{\omega_1\omega_2}^{QS}\right\rangle=\frac{\left(2M_0\right)^2e^{ i\Delta\omega\bar{v}_0}}{\pi^2\sqrt{\bar{\omega}^2-\frac{\Delta\omega^2}{4}}}\int_{0}^{\infty} d\omega'\omega'
{\int\int}_{0}^{+\infty}dxdx'e^{i4M_0\omega'(x-x')}e^{-i4\bar{M}\bar{\omega}\ln(\frac{x'}{x})}e^{-i2\bar{M}\Delta\omega\ln(x'x)}\times
\]
\[
\times e^{-\epsilon(x+x')}e^{-4\Delta M^2\bar{\omega}^2\left[\ln\left(\frac{x'}{x}\right)+\frac{\Delta\omega}{2\bar{\omega}}\ln(x'x)\right]^2},
\]
where we introduced $\Delta\omega=\omega_2-\omega_1$ ,
$\bar{\omega}=\frac{\omega_1+\omega_2}{2}$ and the regulator
$\epsilon$ as before. With the change of variables $x=r\cos(\theta) ,
x'=r\sin(\theta)$ the double integral in $x,x'$ becomes,

\[
\int_0^{\pi/2}d\theta e^{-i4\bar{M}\bar{\omega}\ln\left[\tan(\theta)\right]}
e^{-i2\bar{M}\Delta\omega \ln\left[\sin(\theta)cos(\theta)\right]}
e^{-4\Delta M^2\bar{\omega}^2\left[\ln\left(\tan(\theta)\right)^2+\frac{\Delta\omega}{\bar{\omega}}\ln\left(\tan(\theta)\right)\ln\left(\cos(\theta)\sin(\theta)\right)\right]}
\times
\]
\[\times \int_0^{+\infty}r dr e^{-i4M_0\omega'\left[\sin(\theta)-cos(\theta)\right]r}e^{-i4\bar{M}\Delta\omega\ln(r)}e^{-\epsilon \left[\sin(\theta)+\cos(\theta)\right]r} e^{-8\Delta M^2\bar{\omega}\Delta\omega\ln\left[\tan(\theta)\right]\ln(r)},
\]
where we are using the same $\Delta\omega<<\bar{\omega}$ approximation
used for the study of the classical case in order to simplify the
calculation.

The $r$ integral can be computed using formula (\ref{eq:propiedad_int_gamma}) to obtain

\[
\int_0^{\pi/2}d\theta \Gamma\left(2-8\Delta M^2\bar{\omega}\Delta\omega\ln\left[\tan(\theta)\right]-4\bar{M}\Delta\omega i\right)e^{-i4\bar{M}\bar{\omega}\ln\left[\tan(\theta)\right]}
e^{-i2\bar{M}\Delta\omega \ln\left[\sin(\theta)cos(\theta)\right]}
\times
\]
\[
\times e^{-4\Delta M^2\bar{\omega}^2\left[\ln\left(\tan(\theta)\right)^2+\frac{\Delta\omega}{\bar{\omega}}\ln\left(\tan(\theta)\right)\ln\left(\cos(\theta)\sin(\theta)\right)\right]}
 e^{-\left(2-8\Delta M^2\bar{\omega}\Delta\omega\ln\left[\tan(\theta)\right]-4\bar{M}\Delta\omega i\right)\ln\left(\epsilon+4M_0i\omega'\left[\sin(\theta)-
\cos(\theta)\right]\right)}.
\]

Another change of variable $y=\ln\left(\tan(\theta)\right)$ simplifies
the expression to
\[
-\frac{e^{4\bar{M}\Delta\omega i\ln\left(4M_0\omega'\right)}e^{-2\bar{M}\Delta\omega \pi}}{\left(4M_0\omega'\right)^2}\int_{-\infty}^{+\infty}dy \Gamma\left(2-8\Delta M^2\bar{\omega}\Delta\omega y-4\bar{M}\Delta\omega i\right)e^{-i4\bar{M}\bar{\omega}y}e^{-4\Delta M^2\bar{\omega}^2y^2}\times
\]
\[
\times 
 e^{-\left(2-8\Delta M^2\bar{\omega}\Delta\omega y-4\bar{M}\Delta\omega i\right)\ln\left(\sinh(y/2)-i\epsilon\right)}e^{8\Delta M^2\bar{\omega}\Delta\omega y\ln\left(4M_0\omega'\right)}e^{i4\pi\Delta M^2\bar{\omega}\Delta\omega y}.
\]

Using again the approximation $\Delta\omega<<\bar{\omega}$ the
integral can be further simplified to
\[
-\frac{e^{4\bar{M}\Delta\omega i\ln\left(4M_0\omega'\right)}e^{-2\bar{M}\Delta\omega \pi}}{\left(4M_0\omega'\right)^2}\Gamma\left(2-4\bar{M}\Delta\omega i\right)\int_{-\infty}^{+\infty}dy e^{-i4\bar{M}\bar{\omega}y}e^{-\left(2-4\bar{M}\Delta\omega i\right)\ln\left(\sinh(y/2)-i\epsilon\right)}\times
\]
\[
\times 
 e^{-4\Delta M^2\bar{\omega}^2y^2}e^{8\Delta M^2\bar{\omega}\Delta\omega y\ln\left(4M_0\omega'\right)}.
\]

The last two terms are responsible for the corrections. The Gaussian changes the profile of the number of particles as we discussed before and the other exponential introduces non diagonal terms in the density matrix. Without these terms, the integral in $\omega'$ produces the $\delta(4\bar{M}\Delta\omega)$ dependence seen in Hawking radiation.\\

\section{Summary and outlook}

We have studied the Hawking radiation emitted by a collapsing quantum shell
using the geometric optics approximation. After reviewing the
calculation of the radiation for a classical collapsing null shell, we
proceeded to consider a quantized shell with fluctuating horizons. A
new element we introduce is to take into account the canonically
conjugate variables describing the shell, its mass and the position
along scri minus from which it is incoming. In order to allow
arbitrary superposition of shells with different Schwarzschild radii
the calculation is also performed without assuming from the beginning
that we are considering rays that are close to the horizon.

We find the following results: 

1) Given that we deal with a quantum geometry, the Bogoliubov
coefficients become quantum operators acting on the states of the
geometry. We discover that for computing the Hawking radiation it is
not enough to assume the mean field approximation and consider the
square of the expectation value of the Bogoliubov coefficients
evaluated on the quantum geometry. Such a calculation misleadingly
suggests the Hawking radiation cuts off after a rather short time (the
``scrambling time''). One needs to go beyond mean-field and consider
the expectation value of the square of the Bogoliubov coefficients to
see that the radiation continues forever and that there are departures
from thermality that depend on the initial state of the shell.

2) The resulting Hawking radiation exhibits coherences of the density
matrix, with non vanishing off-diagonal elements for different
frequencies that vanish for the usual calculation on a classical
space-time.  The new correlations that arise in the quantum case have an
imprint of the details of the initial quantum state of the shell. This
indicates that at least part of the information that went into
creating the black hole can be retrieved in the Hawking radiation. It
should be kept in mind that our calculations do not include back
reaction, so to have information retrieval at this level is somewhat
surprising.

3) The non-trivial correlations can be made to vanish taking a shell
with arbitrarily small deviations in the ADM mass. However, such a
shell would have large uncertainties in its initial
position. Therefore such a quantum state would not correspond to a
semi-classical situation. A semi-classical shell will generically have
uncertainty in both the initial position and the ADM mass and will
therefore have non-trivial corrections to the Hawking radiation
through which information can be retrieved.

In our computations we used three simplifying assumptions which should
be improved upon: First, we worked in the geometric optics
approximation which neglects  back-scattering. Moreover, no
back-reaction was considered. This has two implications. On one hand,
information can fall into the black hole and also leak out, violating
no-cloning, in particular the quantum state of the shell is not
modified by the Hawking radiation, which nevertheless gains an imprint
of its characteristics. Moreover, the lack of back reaction eliminates
possible decoherence effects for the shell, which may also lead to information
leakage.  Finally, the collapsing system is a very
simple one: a massless shell. However, the idea that non-trivial
commutation relations between some indicator of the position of the
collapsing system and its ADM mass are expected generically
\cite{carlip} and therefore effects similar to the ones found here are
expected in other collapsing systems.  All in all our calculations
suggests that some level of ``drama at the horizon'' is taking place
that allows to retrieve information from the incoming quantum state.

Summarizing, using the simple example of collapsing quantum shells to
model a fluctuating horizon we have shown that non-trivial quantum
effects can take place, which in particular may allow to retrieve
information from the incoming quantum state at scri plus. A more
careful study is required to determine if the complete information of
the incoming state can be retrieved and if the model generalizes to
more complicated models of horizon formation.

\section*{Appendix 1: Integrals on $I^{-}$ that contribute in the case of a
  quantum black hole} 

The generic expression of interest for the Bogoliubov coefficient
 (\ref{genericexpression}) is,
\[
\left\langle
  \hat{\beta}\right\rangle_{\omega\omega'}=-\frac{\left(4M_{0}\right)^{2}}{2\pi}\sqrt{\frac{\omega'}{\omega}}\underset{\epsilon\to0}{\lim}
{\int_{-\infty}^\infty}dve^{- i\omega'v}{\int_0^\infty\int_0^\infty}dx_{1}dx_{2}\Psi^{*}(4M_{0}x_{1}+v)\times
\]
\[
\times\Psi(4M_{0}x_{2}+v){\int_{-\infty}^\infty}due^{-i\omega u}\underset{I=1,2}{\sum}\psi_{u}^{I}(x_{1})\psi_{u}^{I*}(x_{2})
\]
and the expressions for 
$\psi_{u}^{I}(x)$ are (\ref{eq:autoestados_de_u_1})
and (\ref{eq:autoestados_de_u_2}). Let us show that the integrals,
\[
{\int_0^\epsilon\int_0^\epsilon}dx_{1}dx_{2}+
{\int_0^\epsilon}{\int_\epsilon^1}dx_{1}dx_{2}+{\int_\epsilon^1}{\int_0^\epsilon}dx_{1}dx_{2}
\]
do not contribute in the limit $\epsilon\to0$.
\begin{enumerate}
\item The integral ${\int_0^\epsilon\int_0^\epsilon}dx_{1}dx_{2}$ is
\[
\left\langle
  \hat{\beta}\right\rangle_{\omega\omega'}=-\frac{\left(4M_{0}\right)^{2}}{2\pi}\sqrt{\frac{\omega'}{\omega}}\underset{\epsilon\to0}{\lim}
{\int_{-\infty}^\infty}dve^{- i\omega'v}{\int_0^\epsilon\int_0^\epsilon}dx_{1}dx_{2}\Psi^{*}(4M_{0}x_{1}+v)\times
\]
\[
\times\Psi(4M_{0}x_{2}+v){\int_{-\infty}^\infty}
dve^{- i\omega'v}{\int_0^\epsilon\int_0^\epsilon}dx_{1}dx_{2}\Psi^{*}(4M_{0}x_{1}+v)\times
\]
\[
\times\Psi(4M_{0}x_{2}+v)\frac{1}{4\hbar\left|\ln(\epsilon)\right|}\delta\left(\frac{M_{0}}{\hbar}\frac{x_{1}-x_{2}}{\ln(\epsilon)}-\omega\right)e^{-i\omega v}=
\]
\[
=-\frac{\left(4M_{0}\right)^{2}}{2\pi}\sqrt{\frac{\omega'}{\omega}}\underset{\epsilon\to0}{\lim}{\int_{-\infty}^\infty}dve^{- i\omega'v}{\int_0^\epsilon\int_0^\epsilon}dx_{1}dx_{2}\Psi^{*}(4M_{0}x_{1}+v)\times
\]
\[
\times\Psi(4M_{0}x_{2}+v)\frac{1}{4M_{0}}\delta\left(x_{1}-x_{2}-\frac{\omega\hbar \ln\left(\epsilon\right)}{M_{0}}\right)e^{-i\omega v}.
\]
 This integral vanishes because one can choose 
$\epsilon$ small, in such a way that the argument of the Dirac delta
never vanishes. 
\item The integral ${\int_0^\epsilon}{\int_\epsilon^1}dx_{1}dx_{2}$
is
\[
\left\langle \hat{\beta}\right\rangle_{\omega\omega'}=-\frac{\left(4M_{0}\right)^{2}}{2\pi}\sqrt{\frac{\omega'}{\omega}}\underset{\epsilon\to0}{\lim}{\int_{-\infty}^\infty}dve^{- i\omega'v}{\int_0^\epsilon}{\int_\epsilon^1}dx_{1}dx_{2}\Psi^{*}(4M_{0}x_{1}+v)\times
\]
\[
\times\Psi(4M_{0}x_{2}+v){\int_{-\infty}^\infty}due^{-i\omega
  u}\frac{\exp\left(\frac{iM_{0}}{\hbar}(u-v)\frac{x_{1}}{\ln(\epsilon)}\right)\exp\left(-\frac{iM_{0}}{\hbar}(u-v){\rm
      li}\left(x_{2}\right)\right)}{8\pi\hbar\sqrt{\left|\ln(x)\right|\left|\ln(\epsilon)\right|}}=
\]
\[
=-\frac{\left(4M_{0}\right)^{2}}{2\pi}\sqrt{\frac{\omega'}{\omega}}\underset{\epsilon\to0}{\lim}{\int_{-\infty}^\infty}dve^{-i\omega'v}{\int_0^\epsilon}{\int_\epsilon^1}dx_{1}dx_{2}\Psi^{*}(4M_{0}x_{1}+v)\times
\]
\[
\times\Psi(4M_{0}x_{2}+v)\frac{\delta\left(\frac{M_{0}}{\hbar}\frac{x_{1}-\epsilon}{\ln(\epsilon)}-\frac{M_{0}}{\hbar}\left[{\rm
        li}\left(x_{2}\right)-{\rm li}\left(\epsilon\right)\right]-\omega\right)}{4\hbar\sqrt{\left|\ln(x)\right|\left|\ln(\epsilon)\right|}}e^{-i\omega v}=
\]
\[
=-\frac{\left(4M_{0}\right)^{2}}{2\pi}\sqrt{\frac{\omega'}{\omega}}\underset{\epsilon\to0}{\lim}{\int_{-\infty}^\infty}dve^{-i\omega'v}{\int_0^\epsilon}{\int_\epsilon^1}dx_{1}dx_{2}\Psi^{*}(4M_{0}x_{1}+v)\times
\]
\[
\times\Psi(4M_{0}x_{2}+v)\frac{\delta\left(\frac{x_{1}-\epsilon}{\ln(\epsilon)}-{\rm
      li}\left(x_{2}\right)+{\rm li}\left(\epsilon\right)-\frac{\omega\hbar}{M_{0}}\right)}{4M_{0}\sqrt{\left|\ln(x_{2})\right|\left|\ln(\epsilon)\right|}}e^{-i\omega v}=
\]
\[
=-\frac{4M_{0}}{2\pi}\sqrt{\frac{\omega'}{\omega}}\underset{\epsilon\to0}{\lim}{\int_{-\infty}^\infty}dve^{-i\omega'v}{\int_0^\epsilon}dx_{1}\Psi^{*}(4M_{0}x_{1}+v)\times
\]
\[
\times\Psi(4M_{0}x_{2}\left(x_{1}\right)+v)\sqrt{\frac{\left|\ln(x_{2})\right|}{\left|\ln(\epsilon)\right|}}e^{-i\omega v}
\]
 with
 $x_{2}(x_{1})={\rm li}^{-1}\left(\frac{x_{1}-\epsilon}{\ln(\epsilon)}+{\rm li}\left(\epsilon\right)-\frac{\omega\hbar}{M_{0}}\right)$.
In the integrand 
$\sqrt{\frac{\left|\ln(x_{2})\right|}{\left|\ln(\epsilon)\right|}}$
is bounded above by  $1$ since $x_{2}\in\left(\epsilon,1\right)$ and
$\Psi$ is a wave-packet that we can take to be bounded in all the range
of its variable. Therefore the integral
${\int_0^\epsilon}dx_{1}$
tends to zero when $\epsilon\to0$.
\item The integral ${\int_\epsilon^1}{\int_0^\epsilon}dx_{1}dx_{2}$
yields the same result that
${\int_0^\epsilon}{\int_\epsilon^1}dx_{1}dx_{2}$
since the only change is to substitute $x_{1}$ for $x_{2}$.
\end{enumerate}

\section*{Appendix 2}

Here we present details of the evaluation of the square of the
expectation value of the Bogoliubov coefficients as an approximation
to the number of particles produced. 

If we estimate the expectation value of the number operator using expression (\ref{betasq}) we get,
\[
\left\langle N_{\omega}^{AQS}\right\rangle ={\int_0^\infty}d\omega'\left\langle \hat{\beta}\right\rangle _{\omega\omega'}\left\langle \hat{\beta}\right\rangle _{\omega\omega'}^{*}=\frac{1}{4\pi^{2}}\frac{1}{\omega}\left|\Gamma(1+4\bar{M}\omega i)\right|^{2}e^{-4\pi \bar{M}\omega}{\int_0^\infty}d\omega'\frac{\omega'e^{-\left[\omega+\omega'\right]^{2}\frac{\sigma^{2}}{2}}}{\left(\omega'+\omega\right)^{2}}.\]

Changing variable to $y=\left[\omega+\omega'\right]^{2}\frac{\sigma^{2}}{2}$,
\[
\left\langle N_{\omega}^{AQS}\right\rangle =\frac{\bar{M}}{\pi}\frac{1}{e^{8\bar{M}\pi\omega}-1}{\int_{\frac{\omega^{2}\sigma^{2}}{2}}^\infty}dy\left(y^{-1}-\frac{\omega\sigma}{\sqrt{2}}y^{-3/2}\right)e^{-y}=
\]
\[
=\frac{\bar{M}}{\pi}\frac{1}{e^{8\bar{M}\pi\omega}-1}{\int_{\frac{\omega^{2}\sigma^{2}}{2}}^\infty}dy\frac{e^{-y}}{y}-\frac{\omega\sigma}{\sqrt{2}}{\int_{\frac{\omega^{2}\sigma^{2}}{2}}^\infty}dyy^{-1-1/2}e^{-y}=
\]
\[
=\frac{\bar{M}}{\pi}\frac{1}{e^{8\bar{M}\pi\omega}-1}\left[-\operatorname{Ei}\left(-\frac{\omega^{2}\sigma^{2}}{2}\right)-\frac{\omega\sigma}{\sqrt{2}}\Gamma\left(-\frac{1}{2},\frac{\omega^{2}\sigma^{2}}{2}\right)\right],
\]
where $\operatorname{Ei}$ is the exponential integral and $\Gamma\left(s,x\right)$ is the upper incomplete Gamma function. Taking into account the identities 
$\Gamma\left(s+1,x\right)=s\Gamma\left(s,x\right)+x^{s}e^{-x}$
and $\Gamma\left(\frac{1}{2},x\right)=\sqrt{\pi}\operatorname{erfc}\left(x\right)$, with $\operatorname{erfc}$ the complementary error function,
we get
\begin{equation}
\langle N_\omega^{AQS}\rangle=\frac{\bar{M}}{\pi}\frac{1}{e^{8\bar{M}\pi\omega}-1}\left[-\operatorname{Ei}\left(-\frac{\omega^{2}\sigma^{2}}{2}\right)+2\left\{ \frac{\omega\sigma}{\sqrt{2}}\sqrt{\pi}\operatorname{erfc}\left(\frac{\omega\sigma}{\sqrt{2}}\right)-e^{-\frac{\omega^{2}\sigma^{2}}{2}}\right\} \right],
\end{equation}
which is finite for $\omega\neq0$ and is suppressed as
$e^{-\frac{\omega^2\sigma^2}{2}}$ for $\omega\to+\infty$ (exhibiting
in this approximation a decay that is not present in ordinary thermal
radiation). In fact, the total radiated
energy would be finite since the integral
\[
E={\int_0^\infty}d\omega\hbar\omega\left\langle N_{\omega}^{AQS}\right\rangle,
\]
is convergent.

In the previous calculation we do not have information about the
dependence of intensity of the radiation as a function of time nor its
luminosity, which could be very relevant since the energy loss by the
black hole leads to increased radiation if one were to take into
account back-reaction in the calculations. 

As in the classical case (\ref{casoclasico}) we start by computing the density matrix
\[
\left\langle\rho^{AQS}_{\omega_1,\omega_2}\right\rangle = {\int_0^\infty}d\omega'\left\langle \beta\right\rangle_{\omega_{1}\omega'}\left\langle \beta\right\rangle_{\omega_{2}\omega'}^{*}=\frac{1}{4\pi^{2}\sqrt{\omega_{1}\omega_{2}}}e^{-i\left(\omega_{1}-\omega_{2}\right)\bar{v}_{0}}\Gamma\left(1+4\bar{M}\omega_{1}i\right)\Gamma\left(1-4\bar{M}\omega_{2}i\right)e^{-2\pi \bar{M}\left[\omega_{1}+\omega_{2}\right]}\times
\]
\begin{equation}
\times{\int_0^\infty}d\omega'\frac{\omega'e^{-\left\{ \left[\omega_{1}+\omega'\right]^{2}+\left[\omega_{2}+\omega'\right]^{2}\right\} \frac{\sigma^{2}}{4}}}{\left(\omega'+\omega_{1}\right)\left(\omega'+\omega_{2}\right)}e^{-4\bar{M}i\left[\omega_{1}\ln\left(4M_{0}\left[\omega'+\omega_{1}\right]\right)-\omega_{2}\ln\left(4M_{0}\left[\omega'+\omega_{2}\right]\right)\right]},
\label{eq:approx_density matrix}
\end{equation}
with the same approximation used to compute its diagonal elements (the
number of particles emitted). We assume $\omega_1$ and $\omega_2$ are
close and we expand in $\Delta\omega=\omega_{2}-\omega_{1}\ll\omega_1$
and use $\bar{\omega}=\frac{\omega_{1}+\omega_{2}}{2}$. We obtain,
\[
\left\langle\rho^{AQS}_{\omega_1,\omega_2}\right\rangle=\frac{2\bar{M}}{\pi}\frac{1}{e^{8\bar{M}\pi\bar{\omega}}-1}e^{i\Delta\omega \bar{v}_{0}}{\int_0^\infty}d\omega'\frac{\omega'e^{-\left[\bar{\omega}+\omega'\right]^{2}\frac{\sigma^{2}}{2}}}{\left(\bar{\omega}+\omega'\right)^{2}}e^{4\bar{M}i\Delta\omega\ln\left(4M_{0}\left[\omega'+
\bar{\omega}\right]\right)}+O\left(\Delta\omega\right).\]
 Changing variable to $y=\frac{\left[\bar{\omega}+\omega'\right]^{2}}{\bar{\omega}^{2}}$ we go to
\[
\left\langle\rho^{AQS}_{\omega_1,\omega_2}\right\rangle\sim
\frac{\bar{M}}{\pi}\frac{1}{e^{8\bar{M}\pi\bar{\omega}}-1}e^{i\Delta\omega\bar{v}_{0}}e^{4\bar{M}i\Delta\omega\ln\left(4M_{0}\bar{\omega}\right)}\frac{1}{2}{\int_1^\infty}dy\left(y^{-1}-y^{-3/2}\right)e^{-y\frac{\bar{\omega}^{2}\sigma^{2}}{2}}e^{2\bar{M}i\Delta\omega\ln\left(y\right)}.
\]
Finally,
\[
\left\langle\rho^{AQS}_{\omega_1,\omega_2}\right\rangle\sim\underset{\delta\to0}{\lim}\frac{\bar{M}}{\pi}\frac{1}{e^{8\bar{M}\pi\bar{\omega}}-1}e^{i\Delta\omega\bar{v}_{0}}e^{4\bar{M}i\Delta\omega\ln\left(4M_{0}\bar{\omega}\right)}\frac{1}{2}\times
\]
\[
\times\left[e^{\left(\delta-2\bar{M}i\Delta\omega\right)\ln\left(\frac{\bar{\omega}^{2}\sigma^{2}}{2}\right)}\Gamma\left(-\delta+2\bar{M}i\Delta\omega,\frac{\bar{\omega}^{2}\sigma^{2}}{2}\right)-e^{\left(\frac{1}{2}-2\bar{M}i\Delta\omega\right)\ln\left(\frac{\bar{\omega}^{2}\sigma^{2}}{2}\right)}\Gamma\left(-\frac{1}{2}+2\bar{M}i\Delta\omega,\frac{\bar{\omega}^{2}\sigma^{2}}{2}\right)\right].
\]
The divergent part of the density matrix when $\Delta\omega\to0$ is due to the first term so,
\[
\left\langle\rho^{AQS}_{\omega_1,\omega_2}\right\rangle\sim\underset{\delta\to0}{\lim}\frac{\bar{M}}{2\pi}\frac{1}{e^{8\bar{M}\pi\bar{\omega}}-1}e^{i\Delta\omega\bar{v}_{0}}e^{4Mi\Delta\omega\ln\left(4M_{0}\bar{\omega}\right)}e^{\left(\delta-2\bar{M}i\Delta\omega\right)\ln\left(\frac{\bar{\omega}^{2}\sigma^{2}}{2}\right)}\Gamma\left(-\delta+2\bar{M}i\Delta\omega,\frac{\bar{\omega}^{2}\sigma^{2}}{2}\right).
\]

Now we can calculate the number of particles at time $u_n$ and around $\omega_j$ as
\[
\langle N_{\omega_{j}}^{AQS}\rangle=\frac{1}{\epsilon}
{\int_{j\epsilon}^{\left(j+1\right)\epsilon}\int_{j\epsilon}^{\left(j+1\right)\epsilon}}d\omega_{1}d\omega_{2}e^{-u_n\Delta\omega i}\left\langle\rho^{AQS}_{\omega_1,\omega_2}\right\rangle.
\]
To carry out the integrals we change variables from $\omega_{1,2}$ to
$\Delta\omega$ and $\bar{\omega}$. The result is, 
\[
\langle N_{\omega_{j}}^{AQS}\rangle\sim\frac{\bar{M}}{2\pi}\frac{1}{e^{8\bar{M}\pi\omega_{j}}-1}\underset{\delta\to0}{\lim}{\int_{-\epsilon}^\epsilon}d\left(\Delta\omega\right)\left[1-\frac{\left|\Delta\omega\right|}{\epsilon}\right]e^{\delta\ln\left(\frac{\omega_{j}^{2}\sigma^{2}}{2}\right)}e^{-i\phi\Delta\omega}\Gamma\left(-\delta+2\bar{M}i\Delta\omega,\frac{\omega_{j}^{2}\sigma^{2}}{2}\right),
\]
with $\phi=\left[\left[\frac{2\pi
      n}{\epsilon}\right]-\bar{v}_{0}+4\bar{M}\ln\left(\frac{\omega_{j}\sigma}{\sqrt{2}}\right)-4\bar{M}\ln\left(4M_{0}\omega_{j}\right)\right]$. In order to interpret the result we use an integral representation of
the incomplete Gamma function and reverse the integration order. Then,
\[
\langle
N_{\omega_{j}}^{AQS}\rangle=\frac{\bar{M}}{2\pi}\frac{1}{e^{8\bar{M}\pi\omega_{j}}-1}\underset{\delta\to0}{\lim}e^{\delta\ln\left(\frac{\omega_{j}^{2}\sigma^{2}}{2}\right)}
{\int_{\frac{\omega_{j}^{2}\sigma^{2}}{2}}^\infty}dt\frac{e^{-t}}{t}
{\int_{-\epsilon}^\epsilon}d\left(\Delta\omega\right)\left[1-\frac{\left|\Delta\omega\right|}{\epsilon}\right]e^{-i\Delta\omega\left[\phi-2\bar{M}\ln\left(t\right)\right]}e^{-\delta\ln\left(t\right)}=
\]
\[
=\frac{\bar{M}\epsilon}{\pi}\frac{1}{e^{8\bar{M}\pi\omega_{j}}-1}\underset{\delta\to0}{\lim}e^{\delta\ln\left(\frac{\omega_{j}^{2}\sigma^{2}}{2}\right)}
{\int_{\frac{\omega_{j}^{2}\sigma^{2}}{2}}^\infty}dt\frac{e^{-\left[t+\delta \ln\left(t\right)\right]}}{t}\left\{\frac{\sin\left[\frac{\epsilon}{2}\left(\phi-2\bar{M}\ln\left(t\right)\right)\right]}{\frac{\epsilon}{2}\left(\phi-2\bar{M}\ln\left(t\right)\right)}\right\} ^{2}.
\]
The change of variable 
$y=t/\frac{\omega_{j}^{2}\sigma^{2}}{2}$
clarifies the interpretation of the integral. We get,
\[
\langle N_{\omega_{j}}^{AQS}\rangle=\frac{\bar{M}\epsilon}{\pi}\frac{1}{e^{8\bar{M}\pi\omega_{j}}-1}\underset{\delta\to0}{\lim}{\int_1^\infty}dy\frac{e^{-\frac{\omega_{j}^{2}\sigma^{2}}{2}y}e^{-\delta\ln\left(y\right)}}{y}\left\{\frac{\sin\left[\frac{\epsilon}{2}\left(\alpha-2\bar{M}\ln\left(y\right)\right)\right]}{\frac{\epsilon}{2}\left(\alpha-2\bar{M}\ln\left(y\right)\right)}\right\} ^{2},
\]
where $\alpha$ is the same quantity defined in (\ref{alpha}) with $M$ and $v_0$ replaced by $\bar{M}$ and $\bar{v_0}$. Due to the decreasing exponential we can take the limit in 
$\delta\to0$
getting,
\[
\langle N_{\omega_{j}}^{AQS}\rangle
=\frac{\bar{M}\epsilon}{\pi}\frac{1}{e^{8\bar{M}\pi\omega_{j}}-1}
{\int_1^\infty}dy\frac{e^{-\frac{\omega_{j}^{2}\sigma^{2}}{2}y}}{y}\left\{\frac{\sin\left[\frac{\epsilon}{2}\left(\alpha-2\bar{M}\ln\left(y\right)\right)\right]}{\frac{\epsilon}{2}\left(\alpha-2\bar{M}\ln\left(y\right)\right)}\right\} ^{2}.
\]

 The presence of a factor 
 $\sin^{2}(a)/a^{2}$ 
and the decreasing exponential imply that the integral decreases when 
$\alpha$ grows and also drastically decreases when 
$\alpha<0$. The latter is a result we already knew from the classical
case, but the former is a result of the quantum nature of the black
hole since it is not present if 
$\sigma=0$. Figure (\ref{nwj}) shows the departure from the classical result that appears when one computes the frequency distribution starting from $\left\langle\hat{\beta}\right\rangle$.
We can estimate the time of emission for each frequency using both
extremes. On the one hand the start of the emission happens when 
$\alpha-2\bar{M}\ln\left(y\right)=0$ for $y \sim 1$ that is,
\[
u_{i}-\bar{v}_{0}-4\bar{M}\ln\left(4M_{0}\omega_{j}\right)\sim 0.
\]
We can estimate the end of the emission when 
$\alpha-2 \bar{M}\ln\left(y\right)=0$
for $y\sim\frac{2}{\omega_{j}^{2}\sigma^{2}}$ since larger $y's$ 
are suppressed by the exponential. For $\omega> \sqrt{2}/\sigma$ this value of $y$ is outside the integration range and the total integral is suppressed. For $\omega< \sqrt{2}/\sigma$ we find the condition,
\[
u_{f}-\bar{v}_{0}-4\bar{M}\ln\left(4M_{0}\omega_{j}\right)-2\bar{M}\ln\left(\frac{2}{\omega_{j}^{2}\sigma^{2}}\right)\sim 0,
\]
or,
\[
u_{f}-\bar{v}_{0}-4\bar{M}\ln\left(4M_{0}\sqrt{2}\frac{1}{\sigma}\right)\sim 0.
\]
Note that the time for the end of the emission does not depend on
the frequency. Finally,

\[
\Delta t=u_{f}-u_{i}\sim4\bar{M}\ln\left(4M_{0}\sqrt{2}\frac{1}{\sigma}\right)-4\bar{M}\ln\left(4M_{0}\omega_{j}\right)=-4\bar{M}\ln\left(\frac{\sigma\omega_{j}}{\sqrt{2}}\right).
\]
Restoring the appropriate dimensions,
\begin{equation}
\Delta t\sim-\frac{2R_{s}}{c}\ln\left(\frac{\sqrt{2}\pi\sigma}{\lambda_{j}}\right),
\label{tiempototal}
\end{equation}
where $R_s$ is the Schwarzschild radius and $\lambda_j$ is the
wavelength of frequency $\omega_j$. Recall we are considering frequencies such that  $\omega_j < \sqrt{2}/\sigma$ so that $\Delta t >0$. For $\omega_j >\sqrt{2}/\sigma$ the radiation is suppressed at all times.
One can see that if one integrates $\sum_j\hbar\omega_j\left\langle
  N^{AQS}_{\omega_j}\right\rangle$ with that time interval one obtains
a total emitted energy that is finite. Note that this result
corresponds to a deep quantum regime since we are not considering
$\sigma$ to be very small.

Interestingly, the time (45) corresponds, for the dominant
wavelengths of emission ($R_S$), with the {\em scrambling time}
\cite{harlow}

\begin{equation}
t_{\rm scr} \sim R_s*\ln\left(\frac{R_s}{\ell_{\rm Planck}}\right).
\end{equation}

Quantum information arguments indicate this is precisely the time of information retrieval \cite{1111.6580}.

It should be noted that the result we are obtaining is not due to the
choice of a particular quantum state. To demonstrate this, 
let us now consider a somehow opposite state to the one considered
previously: the case where the shell is in a Gaussian (\ref{eq:gaussiana}) squeezed state with large dispersion in $M$ and small dispersion in $v_0$. The leading quantum correction for such states is obtained by taking the limit $\hbar \to 0$ with
\begin{equation}
 \Delta M = \text{constant} = Z \ell_{\text{Planck}} \; , \; Z \gg 1 \, ; \quad \Delta v_0= \hbar/{\Delta M}.\label{eq:squeezed_state_M_limit}
\end{equation}
In this limit (\ref{eq:coef_bogol_gaussiana}) goes to:

\begin{equation}
\left\langle \hat{\beta}\right\rangle_{\omega\omega'}\to-\frac{2M_{0}}{\pi}\sqrt{\frac{\omega'}{\omega}}e^{-i\left[\omega+\omega'\right]\bar{v}_{0}}{\int_0^\infty}dxe^{i4M_{0}\left[\omega+\omega'\right]x} e^{i 4\bar{M}\omega\ln(x)}e^{-4\Delta M^2\omega^2Ln(x)^2}\label{eq:coef_bogol_gaussiana_dM}
\end{equation}

If we extend the integrand in this expression to $0$ for $x<0$ we recognize the integral as the Fourier transform in $4M_0\left[\omega'+\omega\right]$ of a smooth and rapidly falling function. This implies the Bogoliubov coefficient is a rapidly falling function of $4M_0\left[\omega'+\omega\right]$. It also vanishes for $\omega'=0$ so the total number of emitted particles,

$$\left\langle N_{\omega}^{AQS}\right\rangle ={\int_0^\infty}d\omega'\left\langle \hat{\beta}\right\rangle _{\omega\omega'}\left\langle \hat{\beta}\right\rangle _{\omega\omega'}^{*}$$

is finite for $\omega\neq 0$ as in the previous case.

\section*{Acknowledgement}
We wish to thank Ivan Agull\'o and Don Marolf for discussions. 
This work was supported in part by Grant No. NSF-PHY-1305000,
NSF-PHY-1603630, funds of the Hearne Institute for Theoretical
Physics, CCT-LSU, and Pedeciba.

\end{document}